\documentclass[aps,pre,raggedbottom, nobalancelastpage,
amssymb,twocolumn,groupedaddress,longbibliography]{revtex4}
\usepackage{graphicx}
\usepackage{amsmath}
\usepackage{amsfonts}
\usepackage{amssymb}
\usepackage{epsfig,libertine}
\usepackage[libertine]{newtxmath}
\usepackage{color}
\usepackage{dsfont, hyperref, xcolor}
\usepackage{comment}
\usepackage{enumerate}
\usepackage{mathtools}
\usepackage{physics}
\usepackage{natbib}
\usepackage{svg}
\usepackage{array, dcolumn}
\begin{document}

\title{Keldysh field theory approach to direct 
electric and thermoelectric currents in quantum dots coupled to superconducting leads}

\author{Marco Uguccioni, Luca Dell'Anna}
\affiliation{Dipartimento di Fisica e Astronomia e Sezione INFN, Universit\`{a} degli Studi di Padova, via Marzolo 8, 35131 Padova, Italy}

\date{\today}

\begin{abstract}
We study the transport properties of a quantum dot contacted to two superconducting reservoirs by means of the Keldysh field theory approach, showing how this technique allows us to straightforwardly recover previous results, resulting extremely effective in dealing with quantum transport problem. In particular, we determine the direct current occurring at equilibrium and the electric and thermoelectric currents triggered when the system is driven out of equilibrium by a voltage or a temperature bias, also for a normal-quantum dot-superconductor junction.
The main result of the work is the derivation of the full expression for 
the thermoelectric current in a superconductor-quantum dot-superconductor junction for any values of the temperature difference between the superconducting leads. We show that in the linear response regime, in addition to the Josephson current, a weakly phase-dependent thermoelectric contribution occurs, provided that electron-hole symmetry is broken. Far from linearity, instead, other contributions arise which lead to thermoelectric effects, dominant at weak coupling, also in the presence of particle-hole symmetry.
\end{abstract}

\maketitle

\section{Introduction}
The study of electronic transport in mesoscopic systems has become a cornerstone in condensed matter physics due to its relevance in both fundamental research and technological applications, particularly in quantum computing and nanoelectronics. Quantum dots (QDs), often described as artificial atoms, offer a controllable environment to explore electron transport phenomena thanks to their discrete energy levels and tunable properties. This makes QDs a compelling platform for studying the intricate effects of electronic correlations and transport phenomena under non-equilibrium conditions, providing a direct link between theory and experiment \cite{bib:kouwenhoven}. In this context, the typical transport setup is a QD connected to two normal (N) metallic leads, usually denoted as N-QD-N junction in literature.

Alongside these advancements, the study of superconducting transport in nanoscale devices has also seen significant progress. From a theoretical perspective, with the rise of mesoscopic physics, a deeper understanding of superconducting transport has been developed, focusing on the central concept of Andreev reflection (AR) \cite{bib:andreev,bib:andreev2,bib:beenakker}, namely the process by which an electron in the normal metal is reflected into a hole, with a corresponding creation of an electron pair (Cooper pair) in the superconductor (S). In this context, the Josephson effect \cite{bib:josephson} also plays a crucial role, as it allows the tunneling of Cooper pairs between two superconductors, with the effect of producing a current that flows continuously without any voltage applied, further influencing transport properties at the nanoscale.

When these QDs are coupled to superconducting leads, they create a rich platform for investigating complex interactions among superconductivity, electron correlations, and quantum coherence, especially under non-equilibrium conditions 
{\color{black}\cite{bib:klapwijk, bib:glazman, bib:averin, bib:brutus, bib:levyyeyati2, bib:cuevas2, bib:sun2, bib:eichler, bib:dellanna, bib:koga, bib:yamada, bib:allub}}.
Understanding the mechanisms governing electric and thermoelectric currents in such hybrid systems is essential for advancing nanoscale thermoelectric devices and superconducting quantum technologies. Nowadays, from an experimental point of view, it is possible to produce these systems combining different physical realizations of
QDs connected to superconducting electrodes (for a review see \cite{bib:franceschi}).

{\color {black}Thermoelectric effects are typically suppressed in superconductors due to intrinsic particle-hole symmetry, therefore, detecting them is a challenging task. Breaking particle-hole symmetry usually relies on magnetic fields or impurities, limiting experimental control. Supercondutor-quantum dot-supercondutor (S-QD-S) junctions offer a compelling alternative: the quantum dot's discrete, gate-tunable levels intrinsically break particle-hole symmetry, enabling sizeable and controllable thermoelectric currents, even with ideal superconducting leads. Furthermore, the thermoelectric response can be tuned via the superconducting phase difference, providing direct access to the Andreev bound state spectrum.}

From a theoretical point of view, a powerful and "ab initio" approach to quantum transport in microelectronic devices is provided by the non-equilibrium Green's function technique \cite{bib:jauho}. In particular, in our previous work \cite{bib:uguccioni}, we have shown that by working with a functional integral QFT formalism, also known as Keldysh field theory (KFT) \cite{bib:keldysh,bib:kamenev,bib:diehl}, we can obtain a more transparent, systematic, and efficient means to compute the transport observables (such as currents {\color{black} or noise}) straightforwardly, enabling a comprehensive understanding of transport phenomena in non-equilibrium conditions. 

In this work, we apply the KFT formalism to investigate the direct electric and thermoelectric transport properties in S-QD-S hybrid systems, showing that this formalism still performs very efficiently, providing a proper extension of the theory to superconducting leads. We aim to clarify the interplay between superconducting correlations and non-equilibrium electronic transport 
identifying the key parameters that influence current flow and thermoelectric behaviors.

{\color{black} The goal of the paper is two-fold. We aimed at providing a clear and comprehensive description of the field-theoretical method, already introduced in Ref.~\cite{bib:uguccioni}, here generalized for superconductors, applying it to some paradigmatic examples to benchmark our approach, and deriving the full explicit expression for the thermoelectric currents through a quantum dot contacted to superconducting leads, which is the main and original contribution of our work}.  
The paper is organized as follows: In Sec. \ref{sec:model} we present the system and the physical observable of interest, that is the average current. In Sec. \ref{sec:KFT} we {\color{black} introduce the KFT formalism},
 {\color{black} generalizing what we did for normal leads \cite{bib:uguccioni} to superconducting reservoirs. To test the approach, we} consider several physical cases of interest for which one can obtain exact analyical expressions for the time-independent (direct) currents, and we analyze the results obtained in the paradigmatic case of a single-level non-interacting QD. In particular, in Sec. \ref{sec:JSQDS} we rederive the equilibrium Josephson current in a S-QD-S junction.  
 {\color{black} Sec. \ref{sec:thSQDS} is devoted to study} the thermoelectric effects in a S-QD-S junction due to a temperature bias between the two leads. {\color{black} This section is the most extensive dedicated to the applications of our method and contains novel physical results.
We conclude with a couple of paradigmatic examples where direct currents can be calculate analytically: in Sec. \ref{sec:SQDScoulomb} and Sec. \ref{sec:NQDS} we obtain the non-equilibrium currents in a S-QD-S junction with strong intra-dot interactions and in a N-QD-S junction, respectively, postponing the details in appendixes \ref{app:SQDScoulomb} and \ref{app:NQDS}. 
In these last sections we, therefore, benchmark the results obtained by the KFT formalism with those already reported in the literature, but also calculating the thermoelectric coefficients. 
We, then, give some conclusions in Sec. \ref{sec:conclusion}.}


\section{Model}
\label{sec:model}
We consider a generic interacting quantum dot (QD), describing a confined system with discrete levels, namely 
\begin{equation}
\label{eq:Hdot}
    \hat{H}_{dot}=\sum_\sigma\sum_n \varepsilon_n\hat{d}^\dagger_{n\sigma}\hat{d}_{n\sigma}+\hat{U}(\hat{d}^\dagger_{n\sigma},\hat{d}_{n\sigma}).
\end{equation}
The QD is connected to two superconducting leads (labeled as left and right leads, $a=L, R$), which can be modeled as mean-field Bardeen-Cooper-Schrieffer (BCS) Hamiltonians \cite{bib:BCS} 
\begin{equation}
\begin{split}
\hat{H}_{leads}&=\sum_{a=L,R}\sum_k\bigg[\sum_\sigma\omega_{ka}\hat{c}^\dagger_{k\sigma a}\hat{c}_{k\sigma a}+\\&-\Delta_a\hat{c}^\dagger_{k\uparrow a}\hat{c}^\dagger_{-k\downarrow a}-\Delta^*_a\hat{c}_{-k\downarrow a}\hat{c}_{k\uparrow a}\bigg],
\end{split}
\end{equation}
and the coupling between the dot and the leads is described by the tunneling Hamiltonian 
\begin{equation}
    \hat{H}_{T}=\sum_{a=L,R}\sum_k \sum_n\sum_\sigma\bigg[W_{ka,n}\hat{c}^\dagger_{k\sigma a}\hat{d}_{n\sigma}+W^*_{ka,n}\hat{d}_{n\sigma}^\dagger\hat{c}_{k\sigma a}\bigg].
\end{equation}
The total system is, therefore, described by $\hat{H}=\hat{H}_{leads}+\hat{H}_{dot}+\hat{H}_T$. The operators $\hat{c}^\dagger_{k\sigma a}$ ($\hat{c}_{k\sigma a}$) create (annihilate) electrons with momentum $k$, spin $\sigma=\uparrow,\downarrow$, and dispersion relation $\omega_{ka}$ in the corresponding lead, $a=L,R$.  Since the leads are superconductive, one has to take into account the possibility of electron pairing, that is the formation of Cooper pairs, by introducing complex gap parameters $\Delta_a$. It is important to remark that the superconducting gap parameter is a function of the temperature $T$, i.e. $\Delta_a(T)$, and, according to the BCS theory, the ratio $\Delta_a(T=0)/T_{c}^a$ is a universal quantity and the following asymptotic relation is verified, $\Delta_a(T\lesssim T_c^a)\simeq \sqrt{T_c^a(T-T_c^a)}$, being $T_c^a$ the superconducting critical temperature of the lead $a$ \cite{bib:tinkham}. Similarly, the operators $\hat{d}^\dagger_{n\sigma}$ ($\hat{d}_{n\sigma}$) create (annihilate) electrons with spin $\sigma$ in the discrete states $n=0,1,2,\dots$ of the QD, with non-interacting energy $\varepsilon_n$. Finally, $W_{ka,n}$ are tunneling matrix elements between the leads and the dot. Notice that, in the Hamiltonian $\hat{H}_{dot}$ of the isolated dot, we included arbitrary interactions $\hat{U}$. In this work, we mainly neglect the interaction between electrons in the dot, even though the derivation in the section below is completely general and holds even in the presence of interactions. 
{\color{black} The non-interacting case is reasonable when at least one of the characteristic energies of the problem (i.e. the dot energy, 
the voltage or the coupling between the dot and the reservoirs) 
is large as compared to the dot's interaction energy scale \cite{bib:zamoum}.}

We consider the presence of an external voltage $V$ applied to the system, which creates a current due to a bias in the chemical potentials $\mu_a$ of the leads, $eV=\mu_L-\mu_R$, and we choose the energy origin at the right chemical potential, namely $\mu_R=0$. Moreover we rewrite the gap parameters as $\Delta_a=|\Delta_a|e^{i\phi_a}$, with $\phi_L-\phi_R=\phi$, and we choose the phases as $\phi_L=\phi$ and $\phi_R=0$, without loss of generality. A convenient way to take into account the effects of the external voltage $V$ and the gap phase difference $\phi$ is to perform a \emph{gauge transformation} on the fermionic operators, to move the voltage and the phase to time-dependent hoppings
\begin{equation}
\label{eq:T_gauge}
W_{ka,n}\to W_{ka,n}(t)=W_{ka,n}e^{-i\varphi_a(t)},
\end{equation}
with
\begin{equation}
\varphi_a(t)=\frac{\phi_a}{2}+\mu_at.
\end{equation}
We thus define the current operator from the contact $a=L,R$ to the central region as
\begin{equation}
\begin{split}
\label{eq:Jtransport}
\hat{J}_a(t)&=-e\frac{d}{dt}\hat{N}_a(t)= -e\frac{d}{dt}\sum_k\sum_\sigma \hat{c}^\dagger_{k\sigma a }(t)\hat{c}_{k\sigma a}(t)\\&=-ie\sum_\sigma\big[\hat{H}(t),\hat{c}^\dagger_{k\sigma a}(t)\hat{c}_{k\sigma a}(t)\big]
\end{split}
\end{equation}
{\color{black} 
such that its expectation value can be written as
\begin{equation}
\begin{split}
\label{eq:Jtransport2}
\langle\hat{J}_a(t)\rangle
&=ie\sum_k\sum_n\sum_\sigma \bigg[W_{ka,n}(t)
\langle \hat{c}^\dagger_{k\sigma a}(t)\hat{d}_{n\sigma}(t)\rangle
\\&-W^*_{ka,n}(t)
\langle\hat{d}_{n\sigma}^\dagger(t)\hat{c}_{k\sigma a}(t)\rangle
\bigg],
\end{split}
\end{equation}
where in Eq.~(\ref{eq:Jtransport}) we used the Heisenberg equations of motion, while in Eq.~(\ref{eq:Jtransport2}) we performed the commutation relation explicitly, using the self-consistent definition of the superconducting order parameter, $\Delta_a\sim \sum_k\langle\hat{c}_{-k\downarrow a}\hat{c}_{k\uparrow a} \rangle$, so that $\langle [\hat{H}_{leads},\hat{N}_a]\rangle\propto \sum_k(\Delta_a\langle\hat{c}^\dagger_{k\uparrow a}\hat{c}^\dagger_{-k\downarrow a}\rangle -\Delta^*_a\langle\hat{c}_{-k\downarrow a}\hat{c}_{k\uparrow a} \rangle)=0$. 
}

\section{KFT for BCS Superconductors}
\label{sec:KFT}

In a previous work \cite{bib:uguccioni}, we presented the Keldysh field theory (KFT) formalism as powerful field-theoretical approach useful to describe quantum transport problems driven out of equilibrium. This formalism is based on the Keldysh technique \cite{bib:keldysh}, that is a Green's functions approach which considers the evolution of a system along a closed real time contour $\mathcal{C}$, from $t=-\infty$ to $t=\infty$ and then going back (see Ref.~\cite{bib:kamenev} for a self-contained description). 

Since we want to discuss superconducting transport, first of all we need to generalize the KFT in order to deal with BCS superconductors. The Keldysh action for the two leads can be written in terms of fermionic coherent states, namely Grassmann variables $\bar\chi_{k\sigma a}$, $\bar\chi_{k\sigma a}$, which encode the operators $\hat{c}^\dagger_{k\sigma a}$, $\hat{c}_{k\sigma a}$, respectively, as
\begin{equation}
\begin{split}
    &S_{leads}=\sum_{a=L,R}\sum_k\int_\mathcal{C}dt\; \bigg[\sum_\sigma\bar\chi_{k\sigma a}(t)(i\partial_t-\omega_{ka})\chi_{k\sigma a}(t)+\\&+|\Delta_a|\bar\chi_{k\uparrow a}(t)\bar\chi_{-k\downarrow a}(t)+|\Delta_a|\chi_{-k\downarrow a}(t)\chi_{k\uparrow a}(t)\bigg]\\&=\sum_{a=L,R}\sum_k\int_\mathcal{C}dt\;\begin{pmatrix}
        \bar\chi_{k\uparrow a} \\ \chi_{-k\downarrow a}
    \end{pmatrix}^T_t\begin{pmatrix}
         i\partial_t-\omega_{ka} & |\Delta_a| \\ |\Delta_a| & i\partial_t+\omega_{ka}\end{pmatrix}\begin{pmatrix}
             \chi_{k\uparrow a} \\ \bar\chi_{-k\downarrow a}
         \end{pmatrix}_t.
\end{split}
\end{equation}
In the second line we wrote the action in a quadratic form, where the fermionic vectors, comprising particles and holes degrees of freedoms in single objects, are known as Nambu spinors, while the $2\times 2$ matrix kernel corresponds to the inverse of the so-called Nambu-Gor'kov Green's function \cite{bib:nambu,bib:gorkov}. Following Ref.~{\cite{bib:uguccioni}}, we now split the fermionic fields into forward ($\chi^+$) and backward ($\chi^-$) branches of the contour $\mathcal{C}$ and perform the Keldysh rotation, $\chi^{1(2)}=(\chi^+\pm\chi^-)/\sqrt{2}$ and $\bar\chi^{1(2)}=(\bar\chi^+\mp\bar\chi^-)/\sqrt{2}$, such that an additional Keldysh subspace is also introduced. We, therefore, introduce the Keldysh-Nambu spinors
\begin{equation}
    X_{ka}=\begin{pmatrix} \chi^1_{k\uparrow a} & \Bar{\chi}^1_{-k\downarrow a} & \chi^2_{k\uparrow a} & \Bar{\chi}^2_{-k\downarrow a}\end{pmatrix}^T,
\end{equation}
such that we can rewrite the action for the leads in the following form
\begin{equation}
    S_{leads}=\sum_{a=L,R}\sum_k\int_{-\infty}^\infty dtdt'\;\bar X^T_{ka}(t)\check{g}_{ka}(t-t')X_{ka}(t'),
\end{equation}
where $\check{g}^{-1}_{ka}$ is the inverse Green's function of the isolated leads in the $4\times 4$ Keldysh-Nambu space. More precisely, $\check{g}_{ka}^{-1}$ has the Keldysh structure \cite{bib:kamenev}
\begin{equation}
\label{eq:structure}
    \check{g}^{-1}=\begin{pmatrix}
        [\hat{g}^R]^{-1} & [\hat{g}^{-1}]^K \\ 0 &  [\hat{g}^A]^{-1}
    \end{pmatrix}, \;\;\;\;
    [\hat{g}^{-1}]^K\equiv -[\hat{g}^R]^{-1}\hat{g}^K[\hat{g}^A]^{-1},
\end{equation}
where the retarded ($R$), advanced ($A$) and Keldysh ($K$) components have the $2\times2$ Nambu-Gor'kov structure, namely in frequency space
\begin{equation}
    [\hat{g}_{ka}^{R(A)}]^{-1}(\omega)=\begin{pmatrix}
        \omega-\omega_{ka}\pm i0^+ & |\Delta_a| \\ |\Delta_a| & \omega+\omega_{ka}\pm i0^+
    \end{pmatrix},
\end{equation}
\begin{equation}
    [\hat{g}_{ka}^{-1}]^K(\omega)=\begin{pmatrix}
        2i0^+F_a(\omega) & 0 \\ 0 & 2i0^+F_a(\omega)
    \end{pmatrix}.
\end{equation}
Here, the infinitesimal $\pm i0^+$ are needed to express the physical coupling of the two branches at $t=-\infty$ given by the initial density matrix, while $F_a(\omega)$ is the distribution function for non-interacting fermions in the lead $a$, needed for the correct normalization of the Keldysh partition function, $\mathcal{Z}=1$. Since we are assuming the leads to be at thermal equilibrium with temperatures $T_{a}$, the corresponding distribution functions are given by $F_{a}(\omega)=\tanh[\omega/(2T_{a})]\equiv 1-2f_{a}(\omega)$, being $f_a(\omega)=[e^{\omega/T_a}+1]^{-1}$ the Fermi function. 
Here and in what follows we fix the Boltzmann constant $k_B=1$. To restore it, it will be enough to $T\rightarrow k_BT$ everywhere.  \\
One can thus invert the matrix $\check{g}_{ka}^{-1}$ to get the Keldysh-Nambu-Gor'kov Green's function
\begin{equation}
   \check{g}_{ka}= \begin{pmatrix} \hat{g}^R_{ka} & \hat{g}^K_{ka} \vspace{0.1cm}\\ 0 & \hat{g}^A_{ka} \end{pmatrix}, 
\end{equation}
where, in the Nambu space, 
\begin{eqnarray}
\label{gRAgeneric}
    &&\hat{g}^{R(A)}_{ka}(\omega)=\frac{1}{(\omega\pm i0^+)^2-E^2_{ka}}\begin{pmatrix}
        \omega+\omega_{ka} & -|\Delta_a| \\ -|\Delta_a| & \omega-\omega_{ka}
    \end{pmatrix},
    \\\nonumber\\
\label{eq:FDT}
    &&\hat{g}_{ka}^K(\omega)=F_a(\omega)[\hat{g}_{ka}^R(\omega)-\hat{g}_{ka}^A(\omega)],
\end{eqnarray}
where we introduced the BCS quasi-particle dispersion $E_{ka}=\sqrt{\omega_{ka}^2+|\Delta_a|^2}$. Since the leads are at thermal equilibrium, the $R,A,K$ components are related via the fluctuation-dissipation theorem (FDT) in Eq.~(\ref{eq:FDT}).

Similarly the QD, described by the Grassmann fields $\bar\psi_{n\sigma}$, $\psi_{n\sigma}$ encoding the operators $\hat{d}^\dagger_{n\sigma}$, $\hat{d}_{n\sigma}$, will have the Keldysh action 
\begin{equation}
\begin{split} 
    S_{dot}&=\int_\mathcal{C} dt\; \bigg[\sum_n\sum_\sigma\Bar{\psi}_{n\sigma}(t)(i\partial_t-\varepsilon_n)\psi_{n\sigma}(t)-U(\Bar{\psi}_{n\sigma},\psi_{n\sigma})\bigg]\\&=\sum_n\int_{-\infty}^\infty dtdt'\;\bar\Psi^T_{n}(t)\check{g}^{-1}_{n,0}(t-t')\Psi_{n}(t')+\\&-\int_{-\infty}^\infty dt\;U(\Bar{\psi}^1_{n\sigma},\psi^1_{n\sigma},\Bar{\psi}^2_{n\sigma},\psi^2_{n\sigma}),
\end{split}
\end{equation}
with Keldysh-Nambu spinor 
\begin{equation}
    \Psi_n=\begin{pmatrix} \psi_{n\uparrow}^1 & \bar\psi^1_{n\downarrow} & \psi^2_{n\uparrow} & \bar\psi^2_{n\downarrow}\end{pmatrix}^T,
\end{equation}
and the inverse Keldysh-Nambu-Gor'kov Green's function matrix $\check{g}^{-1}_{n,0}$ for the non-interacting QD in the same form of Eq.~(\ref{eq:structure}) and
\begin{equation}
\label{g0RA}
    [\hat{g}_{n,0}^{R(A)}]^{-1}(\omega)=\begin{pmatrix}
        \omega-\varepsilon_n\pm i0^+ & 0 \\ 0 & \omega+\varepsilon_{n}\pm i0^+
    \end{pmatrix},
\end{equation}
\begin{equation}
    [\hat{g}_{n,0}^{-1}]^K(\omega)=\begin{pmatrix}
        2i0^+F(\omega) & 0 \\ 0 & 2i0^+F(\omega)
    \end{pmatrix}.
\end{equation}
Here, $F(\omega)$ is an unknown distribution function for the non-interacting dot, which will be replaced by finite coupling to the bath later on. 
Finally, the tunneling action reads
\begin{equation}
\begin{split}
    S_{T}&=-\sum_{a=L,R}\sum_k\sum_n\sum_\sigma\int_\mathcal{C} dt\; \bigg[W_{ka,n}(t)\Bar{\chi}_{k\sigma a}(t)\psi_{n\sigma }(t) \\&+ W^*_{ka,n}(t)\Bar{\psi}_{n\sigma}(t)\chi_{k\sigma a}(t)\bigg],\\&=-\sum_{a=L,R}\sum_k\sum_n\int_{-\infty}^\infty dt\; \bigg[\bar X^T_{ka}(t)\check{W}_{ka,n}(t)\Psi_{n}(t) \\&+ \Bar{\Psi}^T_{n\sigma}(t)\check{W}^*_{ka,n}(t)X_{k\sigma a}(t)\bigg],
\end{split}
\end{equation} 
where we introduced the $4\times 4$ time-dependent Keldysh-Nambu matrices
\begin{equation}
\begin{split}
    \check{W}_{ka,n}(t)&=\hat{1} \otimes \hat{\tau}_z\hat{W}_{ka,n}(t)\\&=\begin{pmatrix}
        1 & 0 \\ 0 & 1
    \end{pmatrix} \otimes \begin{pmatrix}
        W_{ka,n}e^{-i\varphi_a(t)} & 0 \\ 0 & -W^*_{ka,n}e^{i\varphi_a(t)},
    \end{pmatrix}
\end{split}
\end{equation}
being $\hat{\tau}_z$ the third Pauli matrix in Nambu space, namely
\begin{equation}
    \hat{\tau}_z=\begin{pmatrix} 
        1 & 0 \\ 0 & -1
    \end{pmatrix},
\end{equation}
and $\varphi_a(t)$ is the time-dependent phase obtained from the gauge transformation reported in Eq.~(\ref{eq:T_gauge}).
 
The Keldysh partition function $\mathcal{Z}$ is normalized by construction \cite{bib:kamenev}
\begin{equation}
\label{normalization}
    \mathcal{Z}=\int D[\bar X,X]D[\bar \Psi,\Psi]e^{iS}\equiv 1,
\end{equation}
where $S=S_{dot}+S_{leads}+S_T$. As previously done in Ref.~\cite{bib:uguccioni}, to make the entire theory meaningful and to compute various observables, we introduce two source fields $A_a$, interacting with the currents
\begin{equation}
\label{eq:Jfields}
    J_a=ie\sum_{k}\sum_n\sum_\sigma\bigg[W_{ka,n}\bar\chi_{k\sigma a}\psi_{n\sigma}-W^*_{ka,n}\bar\psi_{n\sigma}\chi_{ka\sigma}\bigg],
\end{equation}
encoding the {\color{black} relevant operators in Eq.~(\ref{eq:Jtransport2})}, with an action
\begin{equation}
\begin{split}
    S_A&=-\sum_{a=L,R}\int_{\mathcal{C}}dt\; A_{a}(t)J_a(t)\\&=-\sum_{a=L,R}\int^{\infty}_{-\infty}dt\; \bigg[A^+_{a}(t)J^+_a(t)-A^-_{a}(t)J^-_a(t)\bigg]\\&=-2\sum_{a=L,R}\int^{\infty}_{-\infty}dt\; \bigg[A^{cl}_{a}(t)J^q_a(t)-A^q_{a}(t)J^{cl}_a(t)\bigg],
\end{split}
\end{equation}
where we introduced the classical and quantum components of the fields $J^{cl(q)}=(J^+\pm J^-)/2$ and $A^{cl(q)}=(A^+\pm A^-)/2$. Since we want to generate only the physical currents, we can safely set $A_a^{cl}=0$ and thus write the source action in terms of the Keldysh-Nambu spinors $X_{ka}$ and $\Psi_n$ as
\begin{equation}
\begin{split}
    &S_A=-ie\sum_{a=L,R}\sum_k\sum_n\int_{-\infty}^\infty dt\; A^q_a(t)\,\times\\&\bigg[\Bar{X}^T_{k a}(t)\check{\mathcal{W}}_{ka,n}(t)\Psi_{n}(t)-\Bar{\Psi}^T_{n}(t)\check{\mathcal{W}}^*_{ka,n}(t)X_{k a}(t)\bigg],
\end{split}
\end{equation}
with the time-dependent Keldysh-Nambu matrices
\begin{equation}
\begin{split}
    \check{\mathcal{W}}_{ka,n}(t)&=\hat{\tau}_x \otimes \hat{W}_{ka,n}(t)\\&=\begin{pmatrix}
        0 & 1 \\ 1 & 0
    \end{pmatrix} \otimes \begin{pmatrix}
        W_{ka,n}e^{-i\varphi_a(t)} & 0 \\ 0 & W^*_{ka,n}e^{i\varphi_a(t)},
    \end{pmatrix}
\end{split}
\end{equation}
being $\hat{\tau}_x$ the first Pauli matrix in Keldysh space. The expectation values of the currents in Eq.~(\ref{eq:Jfields}) are, then, given by 
\begin{equation}
\label{eq:I_a}
    I_a=\langle J^{cl}_a(t)\rangle=\frac{i}{2}\frac{\delta}{\delta A_a^q(t)}\mathcal{Z}[A_a^q]\bigg|_{A_a^q=0},
\end{equation}
where $\mathcal{Z}[A^q_a]$ is the Keldysh generating functional, given by 
\begin{equation}
    \mathcal{Z}[A^q_a]=\int D[\bar{\chi},\chi]D[\bar{\psi},\psi]e^{iS+iS_A}.
\end{equation}
When the quantum sources are set to zero, namely $A_a^q=0$, one recovers Eq~(\ref{normalization}), namely $\mathcal{Z}=\mathcal{Z}[0]=1$. 

We can now use the fact that the leads are non-interacting to perform the Gaussian integrations over the $\bar{X}_{k a}$, $X_{k a}$ spinors and to find the effective action
\begin{equation}
\begin{split}
\label{eq:s_eff_dot}
    S_{eff}&=S_{dot}-\sum_{a}\sum_k\sum_{n,m}\int_{-\infty}^\infty dt\;dt'\;\Bar{\Psi}^T_{n}(t)[\check{W}^*_{ka,n}(t)+\\&-ieA^q_a(t)\check{\mathcal{W}}^*_{ka,n}(t)]\check{g}_{ka}(t-t')[\check{W}_{ka,m}(t')+\\&+ieA^q_a(t')\check{\mathcal{W}}_{ka,m}(t')]\Psi_{m}(t').
\end{split}
\end{equation}
Notice that, in analogy with what done for the normal case \cite{bib:uguccioni}, one can introduce a self-energy contribution produced by the coupling of the QD with the leads,
\begin{equation}
\label{eq:selfenergycheck}
    \check{\Sigma}_{a,nm}(t,t')=\sum_k \check{W}^*_{ka,n}(t)\check{g}_{ka}(t-t')\check{W}_{ka,m}(t').
\end{equation}
We, then, perform a functional derivative of the effective action over the quantum sources, as in Eq.~(\ref{eq:I_a}), to get the average current (we restore $\hbar$ in the following expressions) 
\begin{equation}
\begin{split}
    I_a(t)&=\frac{e}{2\hbar}\sum_k\sum_{n,m}\int_{-\infty}^\infty dt'\;\mbox{tr}_{K\otimes N}\bigg[\bigg(\check{W}_{ka,n}^{*}\check{g}_{ka}(t,t')\check{\mathcal{W}}_{ka,m}\\&-\check{\mathcal{W}}^*_{ka,n}\check{g}_{ka}(t,t')\check{W}_{ka,m}\bigg)\check{G}_{mn}(t',t)\bigg],
\end{split}
\end{equation}
where $\check{G}_{nm}=-i\langle\Psi_n\Bar{\Psi}^T_m\rangle_{S_{eff}[A^q_a=0]}$ are the components of the exact Keldysh-Nambu-Gor'kov Green's function of the coupled QD, and the subscript $K\otimes N$ means that one has to take the trace over the Keldysh-Nambu space. By performing the trace in the Keldysh space, we get the result for the current
\begin{equation}
\begin{split}
\label{eq:Ia(t)SQD}
    I_a(t)&=\frac{e}{2\hbar}\sum_k\sum_{nm}\int_{-\infty}^\infty dt'\;\mbox{tr}_{N}\bigg[\bigg(\hat{W}^*_{ka,n}\hat{\tau}_z\hat{g}_{ka}^A(t,t')\hat{W}_{ka,m}+\\&-\hat{W}^*_{ka,n}\hat{g}_{ka}^R(t,t')\hat{\tau}_z\hat{W}_{ka,m}\bigg)\hat{G}^K_{mn}(t',t)+\\&+\hat{W}^*_{ka,n}\hat{\tau}_z\hat{g}_{ka}^K(t,t')\hat{W}_{ka,m}\hat{G}^R_{mn}(t',t)+\\&- \hat{W}^*_{ka,n}\hat{g}_{ka}^K(t,t')\hat{\tau}_z\hat{W}_{ka,m}\hat{G}^A_{mn}(t',t)\bigg].
\end{split}
\end{equation}
Notice that we absorbed the time dependence of the hopping matrices into the definition of $\check{g}_{ka}(t,t')$, namely
\begin{equation}
    \hat{g}_{ka}^{p}(t-t')=\begin{pmatrix}
        g^{p 11}_{ka} & g^{p 12}_{ka} 
        \vspace{0.1cm}\\ g^{p 21}_{ka} & g^{p 22}_{ka}
    \end{pmatrix}\,\longrightarrow \,\hat{g}^p_{ka}(t,t')
   \end{equation}
    where 
 \begin{equation}
    \hat{g}^p_{ka}(t,t')=\begin{pmatrix}
        g^{p 11}_{ka}e^{i[\varphi_a(t)-\varphi_a(t')]} & g^{p 12}_{ka}e^{i[\varphi_a(t)+\varphi_a(t')]} 
        \vspace{0.1cm}
        \\ 
        g^{p 21}_{ka}e^{-i[\varphi_a(t)+\varphi_a(t')]} & g^{p 22}_{ka}e^{-i[\varphi_a(t)-\varphi_a(t')]}
    \end{pmatrix},
\end{equation}
with $p=R,A,K$. From the expression above, we see that the off-diagonals terms in the Nambu space produce terms which depend on two times, even in the presence of a time-independent voltage $V\not=0$. This is a peculiar phenomenon, in which a direct voltage in a superconducting junction creates a sinusoidal current, known in literature as AC Josephson effect \cite{bib:josephson}. Thus, in the superconducting case, the general problem does not admit a time-independent solution and the Green's functions depend on two temporal arguments, making the problem much more involved than in the normal case. However, in the following sections, we will focus on some particular cases, for which constant solutions do exist.

The exact Keldysh-Nambu-Gor'kov Green's function of the coupled QD, $\check{G}_{nm}$, can be obtained via the Dyson equation in the Keldysh-Nambu space which, for a non-interacting QD, reads, in frequency space,
\begin{equation}
    \label{eq:Keldysh-Dyson}
\check{G}_{nm}(\omega)=\delta_{nm}\check{g}_{n,0}(\omega)+\sum_{a=L,R}\sum_{l}\check{g}_{n,0}(\omega)\check{\Sigma}_{a,nl}(\omega)\check{G}_{lm}(\omega).
\end{equation}
We see that the form of this Dyson equation consists in three different equations for each component in the Keldysh space, namely retarded, $R$, advanced, $A$, and Keldysh $K$ components, which are, in their turn, $2\times 2$ matrices in Nambu space. 

For later convenience, we also want to compute the local BCS Green's functions in frequency space of the superconductors near the Fermi energy, namely 
\begin{equation}
\begin{split}
\label{eq:gBCS}
    &\hat{g}^{R(A)}_a(\omega)=\sum_{k}\hat{g}^{R(A)}_{ka}(\omega)\\&=\int_{-\infty}^{\infty}\frac{\nu_a(\varepsilon)d\varepsilon}{\varepsilon^2+|\Delta_a|^2-(\omega\pm i0^+)^2}\begin{pmatrix}
     -(\omega\pm i0^+) & |\Delta_a| \\ |\Delta_a| & -(\omega\pm i0^+) \end{pmatrix}
     \\&=\frac{\pi\nu_a}{\sqrt{|\Delta_a|^2-(\omega\pm i0^+)^2}}\begin{pmatrix}
     -(\omega\pm i0^+) & |\Delta_a| \\ |\Delta_a| & -(\omega\pm i0^+) \end{pmatrix},
\end{split}
\end{equation}
where we considered a constant normal density of states (DOS) $\nu_a(\varepsilon)=\sum_k\delta(\varepsilon-\omega_{ka})$, evaluated at the Fermi energy. The local Green's functions above are also known in literature as {quasi-classical Green's functions} \cite{bib:larkin}. The BCS quasi-particle DOS is therefore obtained by taking
\begin{equation}
\label{eq:nuBCS}
    \nu^{BCS}_a(\omega)=\mp\frac{1}{\pi}\Im \mbox{tr } \hat{g}_{a}^{R(A)}(\omega)=\frac{2\nu_a|\omega|}{\sqrt{\omega^2-|\Delta_a|^2}}\theta(|\omega|-|\Delta_a|),
\end{equation}
which is characterized by a typical absence of BCS excitations in the region $-|\Delta_a|<\omega<|\Delta_a|$, that is the reason why  $|\Delta_a|$ is interpreted as an energy gap in the BCS theory. More generally, it is convenient to introduce a DOS matrix in Nambu space, which is
\begin{equation}
\begin{split}
\label{eq:DOSmatrix}
    \hat{\nu}_a(\omega)&=-\frac{1}{2\pi i}[\hat{g}^R_{a}(\omega)-\hat{g}^A_{a}(\omega)]\\&=\frac{\nu_a|\omega|\theta(|\omega|-|\Delta_a|)}{\sqrt{\omega^2-|\Delta_a|^2}}\begin{pmatrix}
        1 & -\frac{|\Delta_a|}{\omega} \\ -\frac{|\Delta_a|}{\omega} & 1
    \end{pmatrix},
\end{split}
\end{equation}
where $\nu^{11}_a=\nu^{22}_a=\nu_a^{BCS}/2$, while $\nu_a^{12}=\nu_a^{21}$ are called {anomalous DOS}.

\section{DC Josephson Currents in S-QD-S Junctions}
\label{sec:JSQDS}
We first consider the case where both the leads are in a superconducting state and no voltage is applied to the system, namely $V=0$. We will see below that we still have a DC current in the junction due to the phase difference $\phi=\phi_L-\phi_R$ of the complex gap parameters, a well known equilibrium phenomenon known as {DC Josephson effect} \cite{bib:josephson}. Here we also consider the two leads at the same temperature $T_L=T_R\equiv T$, and we postpone the discussion to the thermoelectric effects in the next sections. When $V=0$ the Nambu-Gor'kov Green's functions depends only on $t-t'$, therefore we can go in frequency space by a Fourier transform. Moreover, since the total system is now at equilibrium, the Green's functions of the coupled QD satisfies the FDT, namely $\hat{G}^K_{nm}(\omega)=F(\omega)[\hat{G}^R_{nm}(\omega)-\hat{G}^A_{nm}(\omega)]$, which makes simpler the calculations. By using the explicit form of the Nambu matrices $\hat{W}_{ka,n}$, we can perform the trace over the Nambu space in Eq.~(\ref{eq:Ia(t)SQD}), getting the {Josephson DC current}
\begin{equation}
\begin{split}
   I_{a}&=-\frac{e}{h}\sum_{nm}\int_{-\infty}^{\infty}d\omega\; F(\omega)\bigg[\Sigma^{R\,12}_{a,nm}(\omega)G^{R\,21}_{mn}(\omega)+\\&-\Sigma^{A\,12}_{a,nm}(\omega)G^{A\,21}_{mn}(\omega)-\Sigma^{R\,21}_{a,nm}(\omega)G^{R\, 12}_{mn}(\omega)+\\&+\Sigma^{A\,21}_{a,nm}(\omega)G^{A\,12}_{mn}(\omega)\bigg]
 \end{split}
\end{equation}
which can be written as
\begin{equation}
\begin{split}
\label{eq:I^J_a}   
  I_a &=-\frac{2e}{h}\sum_{nm}\int_{-\infty}^{\infty}d\omega\; F(\omega)\Re\bigg[\Sigma^{R\,12}_{a,nm}(\omega)G^{R\,21}_{mn}(\omega)+\\&-\Sigma^{R\,21}_{a,nm}(\omega)G^{R\,12}_{mn}(\omega)\bigg],
\end{split}
\end{equation}
where in the second equality we used the property of the off-diagonal Green's functions in the Nambu formalism, namely $G^{R 12}=[G^{A 21}]^*$, since $[\hat{G}^R]^\dagger=\hat{G}^A$. Here we used the retarded component of the coupling self-energy matrix $\check{\Sigma}_{a,nm}$ in Eq.~(\ref{eq:selfenergycheck}), given by
\begin{equation}
\begin{split}
\label{eq:selfsigmaa,nm}
         &\hat{\Sigma}^{R}_{a,nm}(\omega)=\sum_k 
         {\hat{W}^*_{ka,n}\hat{\tau}_z\,\hat{g}^{R}_{ka}(\omega)\hat{\tau}_z\hat{W}_{ka,m}}\\&\equiv\sum_k\begin{pmatrix}
            	 W^*_{ka,n}g^{R11}_{ka}W_{ka,m} & -W^*_{ka,n}g^{R12}_{ka}e^{i\phi_a}W^*_{ka,m} 
	 \vspace{0.1cm}\\ -W_{ka,n}g^{R12}_{ka}e^{-i\phi_a}W_{ka,m} & W_{ka,n}g^{R22}_{ka}W^*_{ka,m},
         \end{pmatrix},  
\end{split}
\end{equation}
where we consider explicitly the phase of the superconducting order parameter, namely we can consider $\Delta_a$ inside $\hat{g}^R_{ka}$ as real and positive (an amplitude), and therefore $g^{R12}_{ka}=g^{R21}_{ka}$. We can thus symmetrize the current as $I = (I_L - I_R)/2$, to get the total Josephson DC current
\begin{equation}
\begin{split}
 & I=-\frac{e}{h}\sum_{nm}\int_{-\infty}^{\infty}d\omega\; F(\omega)\Re\Big[\Big(\Sigma^{R\,12}_{L,nm}(\omega)
  -\Sigma^{R\,12}_{R,nm}(\omega)\Big) \\&\times \,G^{R\,21}_{mn}(\omega)-\Big(\Sigma^{R\,21}_{L,nm}(\omega)-\Sigma^{R\,21}_{R,nm}(\omega)\Big)G^{R\,12}_{mn}(\omega)\Big].
  \label{IJ}
\end{split}
\end{equation}
It is easy to show that Eq.~(\ref{eq:I^J_a}) and Eq.~(\ref{IJ}) are exactly the same, 
namely $I=I_L=-I_R$, 
since $G^{R\,12}_{mn}\propto (\Sigma^{R\,12}_{L,nm}+\Sigma^{R\,12}_{R,nm})$ and $G^{R\,21}_{mn}\propto (\Sigma^{R\,21}_{L,nm}+\Sigma^{R\,21}_{R,nm})$. Dropping the indices $n,m$ for simplicity, we have
\begin{eqnarray*}
&&\big(\Sigma^{R 12}_{L}-\Sigma^{R 12}_{R}\big)\big(\Sigma^{R 21}_{L}+\Sigma^{R 21}_{R}\big)-
\big(\Sigma^{R 21}_{L}-\Sigma^{R 21}_{R}\big)\big(\Sigma^{R 12}_{L}+\Sigma^{R 12}_{R}\big)\\
&&=2\big(\Sigma^{R 12}_{L}\,\Sigma^{R 21}_{R}-\Sigma^{R 21}_{L}\,\Sigma^{R 12}_{R}\big)\\
&&=2\big[\Sigma^{R 12}_{L}\big(\Sigma^{R 21}_{L}+\Sigma^{R 21}_{R}\big)-
\Sigma^{R 21}_{L}\big(\Sigma^{R 12}_{L}+\Sigma^{R 12}_{R}\big)\big],
\end{eqnarray*}
therefore, the integrands in Eq.~(\ref{IJ}) and Eq.~(\ref{eq:I^J_a}) are the same. 

We now focus on a paradigmatic example, namely a non-interacting dot with a single discrete level $\varepsilon_0$ coupled to two superconductive leads, with real and $k$-independent tunneling matrix elements $W_{L}$ and $W_{R}$, which, without loss of generality, can be taken real. We thus rewrite the current from Eq.~(\ref{IJ}), or equivalently from Eq.~(\ref{eq:I^J_a}) since $I_L=-I_R=I$, simply as follows 
\begin{eqnarray}
\nonumber 
I=\frac{e}{h}\int_{-\infty}^{\infty}d\omega\; F(\omega)\Re\Big[W_L^2 g^{R12}_L(\omega)\\
\times\, \Big(e^{i\phi_L}G^{R\,21}_{00}-e^{-i\phi_L} G^{R\,12}_{00} \Big)\Big]
\end{eqnarray}
where $\hat{g}^R_a(\omega)=\sum_k \hat{g}^R_{ka}(\omega)$. Using the Dyson equation Eq.~(\ref{eq:Keldysh-Dyson}),  we  find, for the retarded Green's function of the dot,
\begin{equation}
\begin{split}
    &G^{R\,11}_{00}(\omega)= \left(\omega+\varepsilon_0-W_L^2g_L^{R22}(\omega)-W_R^2g_R^{R22}(\omega)\right)/D^R(\omega), \\ 
     &G^{R\,22}_{00}(\omega) =  \left(\omega-\varepsilon_0-W_L^2g_L^{R11}(\omega)-W_R^2g_R^{R11}(\omega)\right)/D^R(\omega), \\
     &G^{R\,12}_{00}(\omega)= -\left(W_L^2e^{i\phi_L}g_L^{R12}(\omega)+W_R^2e^{i\phi_R}g_R^{R12}(\omega)\right)/D^R(\omega),\\ 
      &G^{R\,21}_{00}(\omega)=- \left(W_L^2e^{-i\phi_L}g_L^{R12}(\omega)+W_R^2e^{-i\phi_R}g_R^{R12}(\omega)\right)/D^R(\omega),
\end{split}
\end{equation}
and, therefore, 
\begin{equation}
\label{eq:IJdot}
   I=-\frac{4e}{h}W_L^2W_R^2\sin \phi 
   \int_{-\infty}^{\infty}d\omega\; F(\omega)\Im\bigg[\frac{g^{R12}_L(\omega)g^{R12}_R(\omega)}{D^R(\omega)}\bigg].
   \end{equation}
   In terms of the Fermi function, reminding that $F(\omega)=1-2f(\omega)$, Eq.~(\ref{eq:IJdot}) becomes
\begin{equation}
   I=\frac{8e}{h}W_L^2W_R^2\sin\phi\int_{-\infty}^{\infty}d\omega\; f(\omega)\Im\bigg[\frac{g^{R12}_L(\omega)g^{R12}_R(\omega)}{D^R(\omega)}\bigg],
\end{equation}
where $D^R(\omega)=1/\det[\hat{G}^R_{00}]$ and $\phi=\phi_L-\phi_R$.
Using the quasi-classical Green's functions in Eq.~(\ref{eq:gBCS}), one obtains
\begin{equation}
\begin{split}
\label{IJGamma}
    I&=\frac{2e}{h}\Gamma_L\Gamma_R\sin\phi\\&\times \int_{-\infty}^\infty d\omega\; f(\omega)\Im\bigg[\frac{1}{D^R(\omega)}\frac{\Delta^2}{\Delta^2-(\omega+i0^+)^2}\bigg],
\end{split}
\end{equation} 
where, explicitly, 
\begin{equation}
\label{DR}
    \begin{split}
        D^R(\omega)&=\bigg(\omega
        +\frac{1}{2}\frac{(\Gamma_L+\Gamma_R)(\omega+i0^+)}{\sqrt{\Delta^2-(\omega+i0^+)^2}}\bigg)^2-\varepsilon_0^2 \\
        &-\frac{1}{4}\frac{\Delta^2}{\Delta^2-(\omega+i0^+)^2}\,\abs{\Gamma_Le^{i\phi}+\Gamma_R}^2,
\end{split}
\end{equation}
after introducing the couplings $\Gamma_a=2\pi\nu_aW_a^2$ and where we considered the two leads made of the same material, thus $\Delta_L=\Delta_R\equiv \Delta$. One can check that the denominator $D^R(\omega)$ has zeros only for $|\omega|<\Delta$, corresponding to what known in the literature as {Andreev bound states} (ABS) \cite{bib:andreev2}, which are localized states that arise because of the coherent superposition of electron and hole states, confined at the interface between the QD and each of the superconducting leads due to {multiple Andreev reflection} (MAR) processes \cite{bib:andreev}. As we will see in detail below, in such processes the electrons (holes) incident towards the leads are reflected back as holes (electrons). The ABS equation, namely $(\Delta^2-\omega^2)D^R(\omega)=0$, becomes simpler in the symmetric case, $\Gamma_L=\Gamma_R\equiv\Gamma$, when it can be written as
\begin{equation}
    \omega^2\big(\Gamma+\sqrt{\Delta^2-\omega^2}\big)^2- \Gamma^2\Delta^2\cos^2({\phi}/{2})-\varepsilon^2_0\big(\Delta^2-\omega^2\big)=0,
\end{equation}
which, for $\Gamma\gg \Delta$, defining the transmission coefficient $t_0$, 
has solutions at $\omega\simeq \pm \,E_\phi$ where 
\begin{equation}
    E_\phi\equiv \Delta\sqrt{1-t_0\sin^2({\phi}/{2})},\;\;\;\;\;\textrm{with}\;\;\; t_0=\frac{\Gamma^2}{\Gamma^2+\varepsilon_0^2}
\end{equation}
and becomes $\omega\simeq\pm\Delta\cos({\phi}/{2})$ at the Fermi energy $\varepsilon_0=0$. The above results for the DC Josephson current in a single-level QD are perfectly compatible with the literature \cite{bib:levyyeyati}.

In general, the total Josephson current will therefore be a sum of both the contributions from the ABS ($|\omega|<\Delta$) and from the continuous spectrum ($|\omega|>\Delta$). 

By analytic continuation, Eq.~(\ref{IJGamma}) can be written in terms of Matsubara frequencies, $\omega_n=\pi(2 n+1)T$, with $n\in \mathbb{Z}$, as 
\begin{equation}
I=-\frac{e T}{\hbar}\sin\phi\sum_{\omega_n}\frac{1}{D^R(i\omega_n)}\frac{\Gamma^2\Delta^2}{\Delta^2+\omega_n^2}
\end{equation}
which, more explicitly, reads
\begin{equation}
I=\sum_{\omega_n}
\frac{(e/\hbar)\,T\, \Gamma^2\Delta^2\sin\phi}{(\Gamma\Delta\cos\frac{\phi}{2})^2+\omega_n^2(\Gamma+\sqrt{\Delta^2+\omega_n^2})^2+\varepsilon_0^2(\Delta^2+\omega_n^2)}
\end{equation}
useful for a much simpler numerical analysis with respect to Eq.~(\ref{IJGamma}). 
For $\Gamma\gg\Delta$, we get an analytic expression 
\begin{equation}
I\simeq \frac{e\Delta}{2\,\hbar}\sin\phi \frac{t_0\tanh{\big(\frac{\Delta}{2T}\sqrt{1-t_0\sin^2({\phi}/{2})}\big)}}{\sqrt{1-t_0\sin^2({\phi}/{2})}}
\end{equation}
At $T=0$ and $\Gamma\gg \Delta$, the Josephson current reduces to
\begin{equation}
I
\simeq \frac{e\Delta}{2\,\hbar}\frac{t_0\sin\phi}{\sqrt{1-t_0\sin^2({\phi}/{2})}}
=-\frac{2e\Delta}{\hbar}\partial_\phi E_\phi \,.
\end{equation}
At the Fermi energy $\varepsilon_0=0$, at resonance, the transmission coefficient $t_0$ reaches its maximum value, $t_0=1$, and the Josephson current becomes simply
$I\simeq \frac{e\Delta}{2\,\hbar}\frac{\sin\phi}
{\abs{\cos({\phi}/{2})}}$.

\section{Thermoelectric Currents in S-QD-S Junctions}
\label{sec:thSQDS}
In this section we present the last and more technical case for which we have a time-independent current in our system. We compute the thermoelectric current, that is the current generated by a temperature bias between the two leads, in a S-QD-S junction. We thus consider a zero external voltage $V=0$ but keeping the two leads at different temperatures $T_L\not=T_R$. By performing the trace over the Nambu space in Eq.~(\ref{eq:Ia(t)SQD}) and moving to frequency space, we find that the current can be generally divided into two terms, where normal and off-diagonal anomalous parts of the Green's functions contribute, namely $I_a=I_a^{n}+I_a^{o}$, where
\begin{equation}
\begin{split}
\label{normal}
   I^{n}_{a}&=\frac{ie}{h}\sum_{nm}\int_{-\infty}^{\infty}d\omega\; \Gamma^{11}_{a,nm}(\omega)\bigg[G_{mn}^{K\,11}(\omega)+\\&-F_a(\omega)\big(G^{R\,11}_{mn}(\omega)-G^{A\,11}_{mn}(\omega)\big)\bigg],
\end{split}
\end{equation}
has exactly the same form of the spin-symmetric current in Eq.~(\ref{eq:Iqp}) at zero bias, while
\begin{equation}
\begin{split}
\label{anomalous}
   I^{o}_a&=-\frac{e}{h}\sum_{nm}\int_{-\infty}^{\infty}d\omega\; \Re\bigg\{\Sigma^{R\,12}_{a,nm}(\omega)\bigg[G_{mn}^{K\,21}(\omega)+\\&+F_a(\omega)\big(G^{R\,21}_{mn}(\omega)+G^{A\,21}_{mn}(\omega)\big)\bigg]+\\&-\Sigma^{R\,21}_{a,nm}(\omega)\bigg[G_{mn}^{K\,12}(\omega)+F_a(\omega)\big(G^{R\,12}_{mn}(\omega)+G^{A\,12}_{mn}(\omega)\big)\bigg]\bigg\},
\end{split}
\end{equation}
is the contribution which reduces to the Josephson current, Eq.~(\ref{eq:I^J_a}), if the system is at equilibrium, namely when $T_L=T_R$. 

As done in the previous sections, we consider a non-interacting dot with a single energy level $\varepsilon_0$, and we also assume, for simplicity, real and $k$-independent tunneling matrices $W_a$. Without loss of generality, we also consider the phase difference $\phi=\phi_L-\phi_R\equiv \phi_L$ and take $\Delta_L$ and $\Delta_R$ real and positive amplitudes.
Using the Dyson equation in Nambu space for $\hat{G}_{00}^{R(A)}$ and $\hat{G}_{00}^K$, we get
\begin{equation}
\label{eq.hatGK}
\hat G^K_{00}(\omega)=\hat G^R_{00}(\omega) \,\hat \Sigma^K (\omega)\,\hat G^A_{00}(\omega)
\end{equation}
with
\begin{eqnarray}
\label{eq.hatSigmaK}
      && \hspace{-0.75cm}\hat{\Sigma}^K(\omega)=\hat{\Sigma}^K_L(\omega)+\hat{\Sigma}^K_R(\omega)\\
        \nonumber&&=F_L(\omega)\big(\hat{\Sigma}^{R}_L(\omega)-\hat{\Sigma}^{A}_L(\omega)\big)+F_R(\omega)\big(\hat{\Sigma}^{R}_R(\omega)-\hat{\Sigma}^{A}_R(\omega)\big),
\end{eqnarray}    
where, for $p=R,A$, we have
  \begin{equation}      
  \label{eq.Sigma^p}
        \hat{\Sigma}_a^{p}(\omega)=\begin{pmatrix}
            W_a^2g^{p 11}_a (\omega)& -W_a^2e^{i\phi_a}g_a^{p 12} (\omega)
            \vspace{0.1cm}\\ -W_a^2e^{-i\phi_a}g_a^{p 12} (\omega)& W_a^2g^{p 22}_a(\omega)
        \end{pmatrix},
\end{equation}
while
\begin{equation}
\label{eq.GR00}
\begin{split}
G^{p\,11}_{00}(\omega)&=\left(\omega+\varepsilon_0-W_L^2g_L^{p22}(\omega)-W_R^2g_R^{p22}(\omega)\right)/D^p(\omega),\\
G^{p\,22}_{00}(\omega)&=\left(\omega-\varepsilon_0-W_L^2g_L^{p11}(\omega)-W_R^2g_R^{p11}(\omega)\right)/D^p(\omega),\\
G^{p\,12}_{00}(\omega)&= -\left(W_L^2e^{i\phi}g_L^{p12}(\omega)+W^2_Rg_R^{p12}(\omega)\right)/D^p(\omega), \\
G^{p\,21}_{00}(\omega)&=  -\left(W_L^2e^{-i\phi}g_L^{p12}(\omega)+W^2_Rg_R^{p12}(\omega)\right)/D^p(\omega),
\end{split}
\end{equation}
where we dropped the frequency dependence for convenience, and \begin{equation}
\begin{split}
    D^p(\omega)&=\big(\omega-\varepsilon_0-W_L^2g_L^{p11}(\omega)-W_R^2g_R^{p11}(\omega)\big)\\&\times\big(\omega+\varepsilon_0-W_L^2g_L^{p22}(\omega)-W_R^2g_R^{p22}(\omega) \big)\\&-\big(W_L^2e^{i\phi}g_L^{p12}(\omega)+W_R^2g_R^{p12}(\omega)\big)\\
    &\times\big(W_L^2e^{-i\phi}g_L^{p21}(\omega)+W_R^2g_R^{p21}(\omega)\big).
\end{split}
\end{equation}
Differently from the case of a N-QD-S junction, we notice that here the denominator $D^R(\omega)$ has  zeroes corresponding to ABS in the dot. 

Alternatively to Eqs.~(\ref{normal}), (\ref{anomalous}), considering the right term, for simplicity, since $I=I_L=-I_R$, from Eqs.~(\ref{eq:FDT}), (\ref{eq:Ia(t)SQD}) and (\ref{eq.Sigma^p}), we can rewrite the current as follows
\begin{equation}
\begin{split}
I&=-\frac{e}{2h}\int^\infty_{-\infty}d\omega \Tr\big[\left(\hat\Sigma^A_R\tau_z-\tau_z\hat\Sigma^R_R\right)\hat G^K_{00}
\\
&+F_R\big(
(\hat\Sigma^R_R
-\hat\Sigma^A_R)\hat\tau_z\hat G^R_{00}  -\hat\tau_z 
(\hat\Sigma^R_R
-\hat\Sigma^A_R) \hat G^A_{00}\big)
\big]\,,
\end{split}
\end{equation}
where we dropped the $\omega$-dependences to simplify the notation. From Eq.~(\ref{eq.hatGK}) and (\ref{eq.hatSigmaK}) and using once more the Dyson equation, specifically, 
\begin{eqnarray}
&&\hat G^R_{00}\left([\hat{g}_{0,0}^{R}]^{-1}-\hat\Sigma^R_R-\hat\Sigma^R_L\right)=\mathbb{I}\,,\\
&&\left([\hat{g}_{0,0}^{A}]^{-1}-\hat\Sigma^A_R-\hat\Sigma^A_L\right)\hat G^A_{00}=\mathbb{I}\,,
\end{eqnarray}
where $[\hat{g}_{0,0}^{R(A)}]^{-1}$ are given by Eq.~(\ref{g0RA}) whit $n=0$, the single-level index, and introducing 
\begin{eqnarray}
\label{eq.F}
&&{\sf F}=\frac{1}{2}\left(F_R+F_L\right) =1-(f_R+f_L),\\
&&{\sf df} =\frac{1}{2}\left(F_R-F_L\right) =f_L-f_R,
\label{eq.f}
\end{eqnarray}
we can write conveniently the Keldysh component of the Green's function as follows
\begin{equation}
\begin{split}
\label{eq.GK}
\hat G^K_{00}&={\sf F}\left(\hat G^R_{00}-\hat G^A_{00}\right)\\
&+{\sf df} \,\hat G^R_{00}\left[\left(\hat\Sigma^R_R-\hat\Sigma^A_R\right)-\left(\hat\Sigma^R_L-\hat\Sigma^A_L\right)\right]\hat G^A_{00}
\end{split}
\end{equation}
Using Eqs.~(\ref{eq.F})-(\ref{eq.GK}), we can, then, rewrite the current as 
\begin{equation}
\label{IFf}
\begin{split}
I&=\frac{e}{h}\int^\infty_{-\infty}d\omega
\Re\Tr\Big\{{\sf F}
 \big[\hat\tau_z,\hat\Sigma^R_R\big]\hat G^R_{00}-
{\sf df} \,\big(\hat\Sigma^R_R-\hat\Sigma^A_R\big)\hat\tau_z\hat G^R_{00}
\\
&-{\sf df} \;\hat\Sigma^A_R\hat\tau_z\hat G^R_{00}\left[\left(\hat\Sigma^R_R-\hat\Sigma^A_R\right)-\left(\hat\Sigma^R_L-\hat\Sigma^A_L\right)\right]\hat G^A_{00}
\Big\}\,,
\end{split}
\end{equation}
Let us consider all the terms in Eq.~(\ref{IFf}) separately. The first term is the only one which survives for equal temperatures, $T_R=T_L$, since in that case ${\sf df}=0$. It reads
\begin{equation}
I_1=\frac{e}{h}\int^\infty_{-\infty}d\omega\,
{\sf F}(\omega)\Re\Tr\Big\{\big[\hat\tau_z,\hat\Sigma^R_R(\omega)\big]\hat G^R_{00}(\omega)\Big\}
\end{equation}
From Eqs.~(\ref{eq.Sigma^p}) and (\ref{eq.GR00}), it becomes 
\begin{equation}
\label{eq.I1}
\begin{split}
   I_1&=\frac{4e}{h}W_L^2W_R^2\,\sin\phi\int_{-\infty}^{\infty}d\omega\; \big(f_R(\omega)+f_L(\omega)\big)\\
   &\times \,\Im\bigg[\frac{g^{R12}_L(\omega)g^{R12}_R(\omega)}{D^R(\omega)}\bigg],
\end{split}
\end{equation}
which is purely a Josephson current. Since the superconducting gap is temperature dependent, $\Delta(T)$, we have to consider generally $\Delta_R\neq \Delta_L$. The matrix elements of $\hat g^{R(A)}_{R,L}$ are given in Eq.~(\ref{eq:gBCS}) where $\Delta_a$ can be taken real, therefore Eq.~(\ref{eq.I1}) can be written as
\begin{equation}
\label{eq.I1J}
    I_1=\frac{e}{h}\int_{-\infty}^{\infty}d\omega\;\big(f_R(\omega)+f_L(\omega)\big)\,\mathcal{J}(\omega)\sin\phi,
\end{equation}
where $\mathcal{J}(\omega)$ is the Josephson current density given by
\begin{equation}
    \mathcal{J}(\omega)=\Im\bigg[\frac{1}{D^R(\omega)}\frac{\Gamma_L\Delta_L}{\sqrt{\Delta_L^2-(\omega+i0^+)^2}}\frac{\Gamma_R\Delta_R}{\sqrt{\Delta_R^2-(\omega+i0^+)^2}}\bigg],
\end{equation}
where, explicitly, 
\begin{equation}
    \begin{split}
        D^R(\omega)&=\bigg[\omega
        +\frac{1}{2}\Big(\sum_{a}\frac{\Gamma_a\,(\omega+i0^+)}{\sqrt{\Delta_a^2-(\omega+i0^+)^2}}\Big)\bigg]^2 
         -\varepsilon_0^2\\
        &-\frac{1}{4}\Big(\sum_{a}\frac{\Gamma_a\Delta_a e^{i\phi_a}}{\sqrt{\Delta_a^2-(\omega+i0^+)^2}}\Big)
        \Big(\sum_{a}\frac{\Gamma_a\Delta_a e^{-i\phi_a}}{\sqrt{\Delta_a^2-(\omega+i0^+)^2}}\Big)    
\end{split}
\end{equation}
Also in this case, by analytic continuation, Eq.~(\ref{eq.I1J}) can be written in terms of Matsubara frequencies, $\omega_n=\pi (2 n+1)T_L$ and $\omega^\prime_n=\pi (2 n+1)T_R$, with $n\in \mathbb{Z}$, as 
\begin{equation}
I_1=-\frac{e}{2\hbar} \,\bigg(T_L\sum_{\omega_{n}}\mathcal{J}(i\omega_{n})+T_R\sum_{\omega^\prime_{n}}\mathcal{J}(i\omega'_{n})\bigg)\sin\phi\,.
\end{equation}
The second contribution is given by
\begin{equation}
I_2=\frac{e}{h}\int^\infty_{-\infty}d\omega\, {\sf df}(\omega)
\Re\Tr\Big[\big(\hat\Sigma^A_R(\omega)-\hat\Sigma^R_R(\omega)\big)\hat\tau_z\hat G^R_{00}(\omega)\Big]
\end{equation}
Let us define for convenience, from Eq.~(\ref{eq:DOSmatrix}), the matrix
\begin{equation}
\hat\rho_a(\omega)=\frac{1}{\nu_a}
\begin{pmatrix}
        \nu^{11}_a(\omega) & \nu^{12}_a(\omega)e^{i\phi_a}  \vspace{0.1cm}\\ \nu^{21}_a(\omega)e^{-i\phi_a}  & \nu^{22}_a(\omega) 
    \end{pmatrix}
\end{equation}
so that we can write
\begin{equation}
\label{Sigma_rho}
\hat\Sigma^R_a(\omega)-\hat\Sigma^A_a(\omega)=
-i\,\Gamma_a \begin{pmatrix}
\rho^{11}_a(\omega)&-\rho^{12}_a(\omega) \vspace{0.1cm}\\
-\rho^{21}_a(\omega)&\rho^{22}_a(\omega)
\end{pmatrix}
\end{equation}
where $\rho_a^{11}=\rho_a^{22}=\frac{|\omega|\theta(|\omega|-\Delta_a)}{\sqrt{\omega^2-\Delta_a^2}}$ and $\rho_a^{12}={\rho_a^{21*}}=-\frac{\Delta_a}{\omega}\rho_a^{11}e^{i\phi_a}$.
In our case, for $a=R$, $\rho_a^{11}$ and $\rho_a^{12}$ are both real since we chose $\phi_R=0$.  
We can, then, rewrite $I_2$ splitting it in two parts, $I_2=I_{2{\rm a}}+I_{2{\rm b}}$, as follows
\begin{equation}
\label{eq:I2aI2b}
\begin{split}
I_{2{\rm a}}=\frac{e}{h}\Gamma_R \int^\infty_{-\infty}\hspace{-0.1cm}
d\omega\,{\sf df}(\omega) \,\rho_R^{11}(\omega)
\Im\big(G^{R22}_{00}(\omega)-G^{R11}_{00}(\omega)\big)\\
I_{2{\rm b}}=\frac{e}{h}\Gamma_R\int^\infty_{-\infty}\hspace{-0.1cm}
d\omega\,{\sf df}(\omega) \,\rho_R^{12}(\omega)\Im\big(G^{R 12}_{00}(\omega)-G^{R 21}_{00}(\omega)\big)
\end{split}
\end{equation}
Quite in general terms in the integral we would have had $\Im\big(\rho_R^{11}G^{R 11}_{00}-\rho_R^{22}G^{R 22}_{00}\big)$ and $\Im\big(\rho_R^{21}G^{R 12}_{00}-\rho_R^{12}G^{R 21}_{00}\big)$. 
From Eq.~(\ref{eq.GR00}), Eq.~(\ref{eq:I2aI2b}) can be written as
\begin{equation}
\begin{split}
\label{I2ab}
\hspace{-0.2cm} 
I_{2{\rm a}}=&-\frac{2 e}{h}\Gamma_R\int^\infty_{-\infty}\hspace{-0.1cm}
d\omega\,{\sf df}(\omega)\,
\rho^{11}_R(\omega)
\Im\Big(\frac{\varepsilon_0}{D^R(\omega)}\Big)\\
\hspace{-0.2cm} 
I_{2{\rm b}}=&-\frac{e\,\Gamma_R \Gamma_L}{h\pi\nu_L}\int^\infty_{-\infty}\hspace{-0.1cm}
d\omega\,{\sf df}(\omega)\,
\rho^{12}_R(\omega)
\Re\Big(\frac{g_L^{R12}(\omega)}{D^R(\omega)}
\Big)\sin\phi
\end{split}
\end{equation}
We notice that $I_{2{\rm a}}$ is manifestly an odd function of $\varepsilon_0$, namely $I_{2a}(\varepsilon_0)=-I_{2a}(-\varepsilon_0)$, and since $G^{R 22}_{00}(-\varepsilon_0)=G^{R 11}_{00}(\varepsilon_0)$, 
it can also be written as
\begin{equation}
I_{2{\rm a}}=-\frac{2 e}{h}\Gamma_R\int^\infty_{-\infty}\hspace{-0.1cm}
d\omega\,{\sf df}(\omega) \,\rho_R^{11}(\omega)\Im\big(G^{R11}_{00}(\omega)\big)\\
\end{equation}
in agreement with Ref.~\cite{bib:kleeorin}. As we will see, $I_{2{\rm a}}$ gives the dominant contribution to the thermoelectric current, allowing us to define the thermoelectric coefficient in the linear response.\\
The current $I_{2{\rm b}}$, instead, can be seen as a thermoelectric correction to the Josephson current. 

The last term in Eq.~(\ref{IFf}) is 
\begin{equation}
\label{IFflast}
\begin{split}
\hspace{-0.05cm}I_3&=\frac{e}{h}\int^\infty_{-\infty}\hspace{-0.05cm}
d\omega \,{\sf df}(\omega) 
\Re\Tr\Big\{\hat\Sigma^A_R(\omega)\hat\tau_z\hat G^R_{00}(\omega)\\
&\times \left[\left(\hat\Sigma^R_L(\omega)-\hat\Sigma^A_L(\omega)\right)-\left(\hat\Sigma^R_R(\omega)-\hat\Sigma^A_R(\omega)\right)\right]\hat G^A_{00}(\omega)
\Big\},
\end{split}
\end{equation}
and, using Eq.~(\ref{Sigma_rho}), it can be written as
\begin{equation}
\label{IFflast2}
\begin{split}
\hspace{-0.08cm}I_3&=\frac{e}{h}\int^\infty_{-\infty}\hspace{-0.05cm}
d\omega \,{\sf df}(\omega) 
\Im\Big\{
\sum_{ijk\ell}\Big[\big(\Gamma_R \rho^{ij}_R(\omega)-\Gamma_L\rho^{ij}_L(\omega)\big)\\
&\times (-1)^{(\ell+i+j)}G^{A jk}_{00}(\omega)\,\Sigma^{A k\ell}_R (\omega)\,G^{R \ell i}_{00}(\omega)
\Big]
\Big\}.
\end{split}
\end{equation}
Using the analytical properties of the Greens' functions and $\rho^{11}_a=\rho^{22}_a$ and $\rho^{12}_a=\rho^{21*}_a$, we can halve the number of terms summing them in pairs such that Eq.~(\ref{IFflast2}) simplifies as 
\begin{equation}
\label{IFflast2_2}
\begin{split}
\hspace{-0.08cm}I_3&=\frac{2e}{h}\int^\infty_{-\infty}\hspace{-0.05cm}
d\omega \,{\sf df}(\omega) 
\Im\Big\{
\sum_{ik\ell}\Big[\big(\Gamma_R \rho^{1i}_R(\omega)-\Gamma_L\rho^{1i}_L(\omega)\big)\\
&\times (-1)^{\ell+i}\,G^{A ik}_{00}(\omega)\,\Sigma^{A k\ell}_R (\omega)\,G^{R \ell 1}_{00}(\omega)
\Big]
\Big\}.
\end{split}
\end{equation}
As we will see, in the symmetric case, $\Gamma_R=\Gamma_L$, this last term, $I_3$, for small $\phi$, is strongly suppressed by a small $\delta T$, due to the coexistence of both the term $(f_L-f_R)$ and the term $(\hat\rho_R-\hat\rho_L)$.

In conclusions, the full current is, then, given by
\begin{equation}
\label{eq:totcurr}
I=I_1+I_{2\rm a}+I_{2\rm b}+I_3
\end{equation}
where $I_1$ is reported in Eq.~(\ref{eq.I1J}), $I_{2\rm a}$ and $I_{2\rm b}$ in Eqs.~(\ref{I2ab}) and $I_3$ in Eq.~(\ref{IFflast2_2}).

\subsubsection{Linear response}
\begin{figure}
  \includegraphics[width=\linewidth]{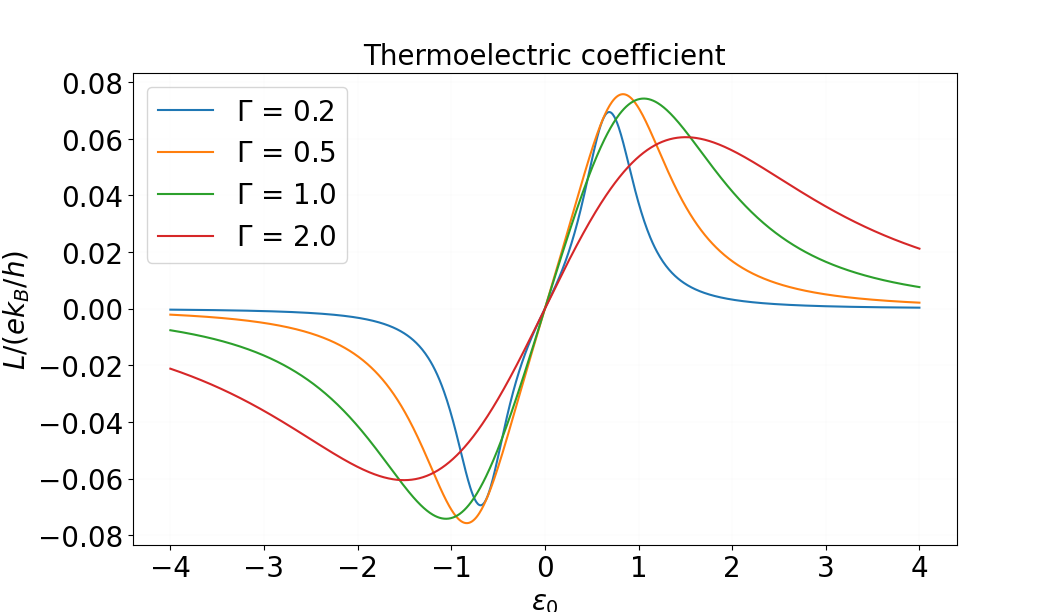}
  \caption{Thermoelectric coefficient $L$ in units of $ek_B/h$ as a function of the dot's energy $\varepsilon_0$ in a S-QD-S junction with energy gap $\Delta=0.5$ and temperature $\beta=10$, for different values of the coupling $\Gamma$.}
  \label{fig:7}
\end{figure}

In the linear response regime, namely at small $\delta T=T_L-T_R$ and small $\phi$, for a symmetric junction with $\Gamma^L=\Gamma^R\equiv \Gamma$ and $\Delta_L=\Delta_R\equiv\Delta$, we can consider $\Delta(T_L)\simeq \Delta(T_R)$. Therefore the analytical expression for the total current becomes much simpler, since only two independent terms coming from $I_1$ and $I_{2{\rm a}}$ contribute, since $I_{2{\rm b}}$ and $I_3$ are subleading terms, $I_{2{\rm b}}\sim O(\phi \delta T)$ and $I_3\sim O(\phi\delta T)+O({\delta T}^2)$, 
\begin{equation}
I 
\simeq J\phi+L\,\delta T
\end{equation}
where, from Eq.~(\ref{eq.I1J}), 
\begin{equation}
\begin{split}
    J&=\frac{2e}{h}\int_{-\infty}^\infty d\omega\; f(\omega)\Im\Big(\frac{1}{D^R(\omega)}\frac{\Gamma^2\Delta^2}{\Delta^2-(\omega+i0^+)^2}\Big)\Big|_{\phi=0},
\end{split}
\end{equation}
is the Josephson coefficient, while, from $I_{2{\rm a}}$ in Eq.~(\ref{I2ab}), 
\begin{equation}
\begin{split}
    L=-\frac{2e}{h} \varepsilon_0\Gamma \int_{-\infty}^{\infty} \hspace{-0.05cm} d\omega
    \frac{df}{dT}(\omega)\,\rho(\omega)
    \Im\Big(\frac{1}{D^R(\omega)}\Big)\Big|_{\phi=0},
\end{split}
\end{equation}
with $\rho(\omega)=\frac{|\omega|\theta(|\omega|-\Delta)}{\sqrt{\omega^2-\Delta^2}}$ and $D^R(\omega)$ given by Eq.~(\ref{DR}) with $\Gamma_R=\Gamma_L$, namely 
$D^R(\omega)=\big[\big(\omega+\frac{\Gamma(\omega+i0^+)}{\sqrt{\Delta^2-(\omega+i0^+)^2}}\big)^2-\frac{\Gamma^2\Delta^2\cos^2\frac{\phi}{2}}{\Delta^2-(\omega+i0^+)^2}-\varepsilon_0^2\big] $, 
is the thermoelectric coefficient  due to the quasi-particle transmission processes. 
We notice that ABS do not contribute to the thermoelectric transport due to the theta-term, which is non-zero only outside the the gap, i.e. $|\omega|>\Delta$. In Fig.~\ref{fig:7} we show the thermoelectric coefficient as a function of $\varepsilon_0$ for different values of the coupling $\Gamma$, at fixed $\Delta=0.5$ and $\beta=10$. Also here we observe the typical antisymmetric behavior of $L$ with respect to $\varepsilon_0$, as discussed in the other cases above and for normal leads \cite{bib:uguccioni}. We stress again the fact that the thermoelectric effects in a QD connected to superconducting leads are weaker with respect to the case of N-QD-N junctions, because a part of thermal energy is required for breaking Cooper pairs before transmitting single quasi-particles.

For arbitrary values of $\phi$ we have to consider also $I_{2\rm b}$ and $I_3$ since both have terms $\sim O(\delta T)$ for finite values of $\phi$. However, as we will see for a quite general exact symmetric case, there is a perfect cancellation of $I_{2\rm b}$ with $I_3$, therefore, for any values of $\phi$ and up to $(\delta T)^2$ we have
\begin{equation}
\label{current_LR}
I= I_1+I_{2\rm a}+{O}(\delta T^2)
\end{equation}
where $I_1$ is given by Eq.~(\ref{eq.I1J}) and $I_{2\rm a}$ is in Eq.~(\ref{I2ab}), with equal couplings $\Gamma_R=\Gamma_L$ and $\Delta_L-\Delta_R\simeq \frac{\partial\Delta}{\partial T} \delta T$, $f_L-f_R\simeq \frac{\partial f}{\partial T} \delta T$.

\subsubsection{Symmetric case}
For a perfect symmetric case, namely for $\Gamma=\Gamma_R=\Gamma_L$,  $\nu_L=\nu_R$, and $\Delta_R=\Delta_L$ (a condition which can be approximately obtained using same superconducting leads in the limit of very small $\delta T$, in the linear response regime, or using two different superconductors at different temperatures providing that $\Delta_R(T_R)=\Delta_L(T_L)$, valid also far from linear response) we have that the total current is exactly given by 
\begin{equation}
\label{eq:totcurr_symm}
I= I_1+I_{2\rm a}\,.
\end{equation}
The proof comes in what follows. In the symmetric case we have 
$\rho^{11}_R=\rho^{11}_L$, therefore, the terms with $i=1$ in Eq.~(\ref{IFflast2_2}) vanish. Moreover, for $i=2$ the term with $k\neq \ell$ is an odd function of $\omega$. 
As a result, since in the perfect symmetric case we have $\rho^{12}_L=e^{i\phi}\rho^{12}_R$, Eq.~(\ref{IFflast2_2}) reduces to
\begin{equation}
\label{IFflast2_equalGamma}
\begin{split}
\hspace{-0.08cm}I_3&=\frac{2e}{h}\Gamma\int^\infty_{-\infty}\hspace{-0.05cm}
d\omega \,{\sf df}(\omega) \,\rho^{12}_R(\omega)
\Im\Big[
\big(1-e^{i\phi}\big) \Sigma^{A 11}_R(\omega)\\
&\times 
\Big(G^{A 22}_{00}(\omega)\,G^{R 2 1}_{00}(\omega)-G^{A 21}_{00}(\omega)\,G^{R 1 1}_{00}(\omega)\Big)\Big],
\end{split}
\end{equation}
where we used $\Sigma^{p11}_a=\Sigma^{p22}_a$. 
Using the following equations
\begin{eqnarray}
&&G^{p21}(\omega)=- (1+e^{-i\phi})\frac{\Gamma}{2\pi\nu}\,\frac{g^{p12}(\omega)}{D^p(\omega)}\\
&&g^{A ij}(\omega)\,\theta(|\omega|-\Delta)=-g^{R ij}(\omega)\,\theta(|\omega|-\Delta)
\label{eq.g_theta}
\end{eqnarray}
where the latter means that $g^{pij}(\omega)$ appearing under the integral are purely imaginary since $\abs{\omega}\ge\Delta$, and noticing that
\begin{equation}
(1-e^{i\phi})(1+e^{-i\phi})=-2i\sin\phi
\end{equation}
we can rewrite Eq.~(\ref{IFflast2_equalGamma}) as follows
\begin{equation}
\label{IFflast2_equalGamma2}
\begin{split}
\hspace{-0.08cm}I_3&=-\frac{2e\Gamma^2}{h\pi \nu}\int^\infty_{-\infty}\hspace{-0.05cm}
d\omega \,{\sf df}(\omega) \,\rho^{12}_R(\omega)
\Re\bigg[
\Sigma^{A 11}_R(\omega)\\
&\times g^{R12}(\omega)
\bigg(\frac{G^{A 22}_{00}(\omega)}{D^R(\omega)}+\frac{G^{R 11}_{00}(\omega)}{D^A(\omega)}\bigg)\bigg]\sin\phi.
\end{split}
\end{equation}
Using Eqs.~(\ref{eq.GR00}) and (\ref{eq.g_theta}), under the integral, we have
\begin{equation}
\label{GD+GD}
\bigg(\frac{G^{A 22}_{00}(\omega)}{D^R(\omega)}+\frac{G^{R 11}_{00}(\omega)}{D^A(\omega)}\bigg)\theta(|\omega|-\Delta)=\frac{2\omega\,\theta(|\omega|-\Delta)}{\abs{D^R(\omega)}^2}
\end{equation}
Now, since $\Sigma^{p11}_R=\Sigma^{p22}_L$, we realize that
\begin{equation}
\label{omegaSigma}
\omega \,\Sigma^{A11}_R(\omega)\,\theta(|\omega|-\Delta)=\frac{1}{8}\big(D^R(\omega)-D^A(\omega)\big)\theta(|\omega|-\Delta)
\end{equation}
Replacing Eq.~(\ref{GD+GD}) in Eq.~(\ref{IFflast2_equalGamma2}) and using Eq.~(\ref{omegaSigma}) we get
\begin{equation}
\label{IFflast2_equalGamma3}
\begin{split}
\hspace{-0.08cm}I_3&=-\frac{e\Gamma^2}{2h\pi \nu}\int^\infty_{-\infty}\hspace{-0.05cm}
d\omega \,{\sf df}(\omega) \,\rho^{12}_R(\omega)
\Re\bigg[g^{R12}(\omega)
\\
&\times 
\bigg(\frac{1}{D^A(\omega)}-\frac{1}{D^R(\omega)}\bigg)\bigg]\sin\phi.
\end{split}
\end{equation}
Since in the domain of integration ($\abs{\omega}\ge \Delta$) the term $g^{R12}(\omega)$ is purely imaginary, we can write at the end
\begin{equation}
\label{IFflast2_equalGamma3}
\begin{split}
\hspace{-0.08cm}I_3&=\frac{e\Gamma^2}{h\pi \nu}\int^\infty_{-\infty}\hspace{-0.05cm}
d\omega \,{\sf df}(\omega) \,\rho^{12}_R(\omega)
\Re\bigg(\frac{g^{R12}(\omega)}{D^R(\omega)}\bigg)\sin\phi\,.
\end{split}
\end{equation}
We showed, therefore, that for the symmetric case, we have
\begin{equation}
I_3=-I_{2\rm b}
\end{equation}
as one can see from Eq.~(\ref{I2ab}). As a result, the last two terms in Eq.~(\ref{eq:totcurr}) cancel out completely, getting Eq.~(\ref{eq:totcurr_symm}).

\subsubsection{Results beyond linear response}

\begin{figure}[h!]
 \includegraphics[width=0.8\linewidth]{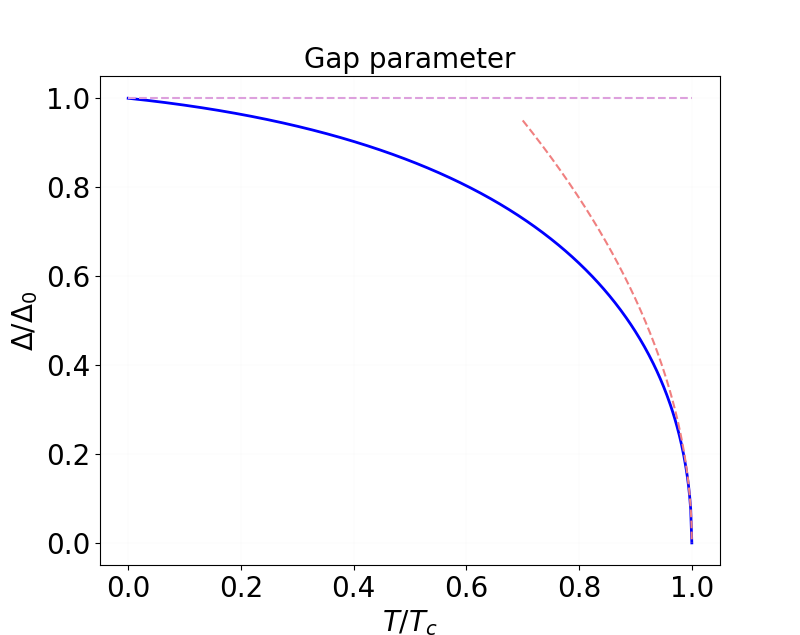}
  \caption{Gap parameter $\Delta(T)$, in units of $\Delta_0\equiv\Delta(T=0)$, as a function of the temperature $T$, in units of $T_c$, according to the approximated expression given by Eq.~(\ref{deltaT}) with $\eta=1$. The dashed lines are the maximum value, Eq.~(\ref{delta0}), and the asymptotic critical behavior, Eq.~(\ref{deltac}).}
  \label{fig:delta}
\end{figure}

\begin{figure}[h!]
  \includegraphics[width=0.95\linewidth]{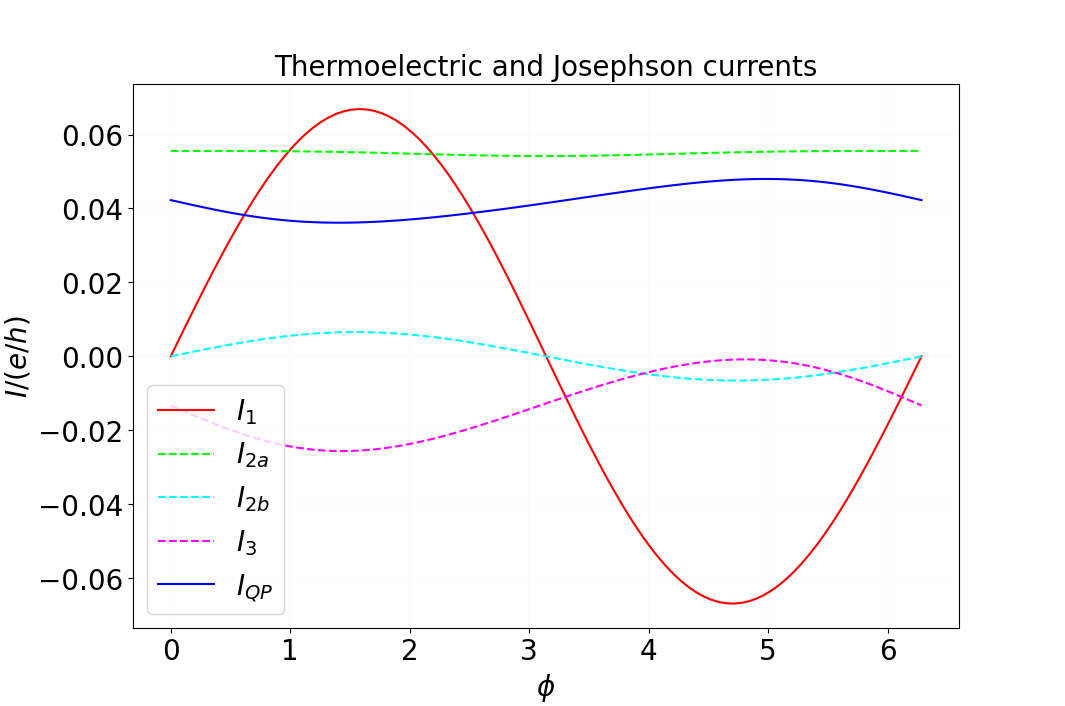}
  \caption{Currents, in units of $e/h$, for $\Gamma_R=\Gamma_L=0.3$, {$\varepsilon_0=1$}, $T_R=0.2$, $T_L=0.4$, $\Delta_0=1$ (from Eq.~(\ref{deltaT}), $\Delta_R\approx 0.92$, $\Delta_L\approx 0.72$), as functions of $\phi$.}
  \label{fig:currents1}
\end{figure}

\begin{figure}[h!]
  \includegraphics[width=\linewidth]{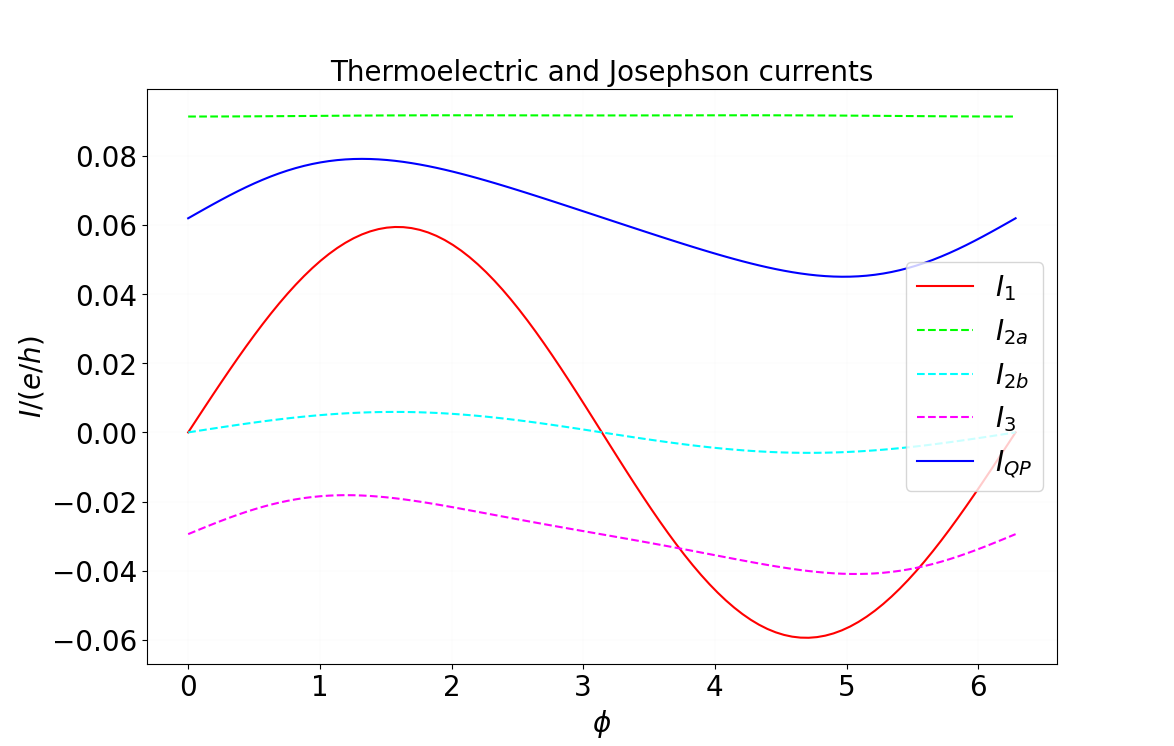}
  \caption{Currents, in units of $e/h$, for $\Gamma_R=\Gamma_L=0.3$,  $\varepsilon_0=1$, $T_R=0.1$, $T_L=0.5$, $\Delta_0=1$ (from Eq.~(\ref{deltaT}), $\Delta_R\approx 0.97$, $\Delta_L\approx 0.51$), as functions of $\phi$.}
  \label{fig:currents_phi}
  \end{figure}

\begin{figure}[h!]
    \includegraphics[width=\linewidth]{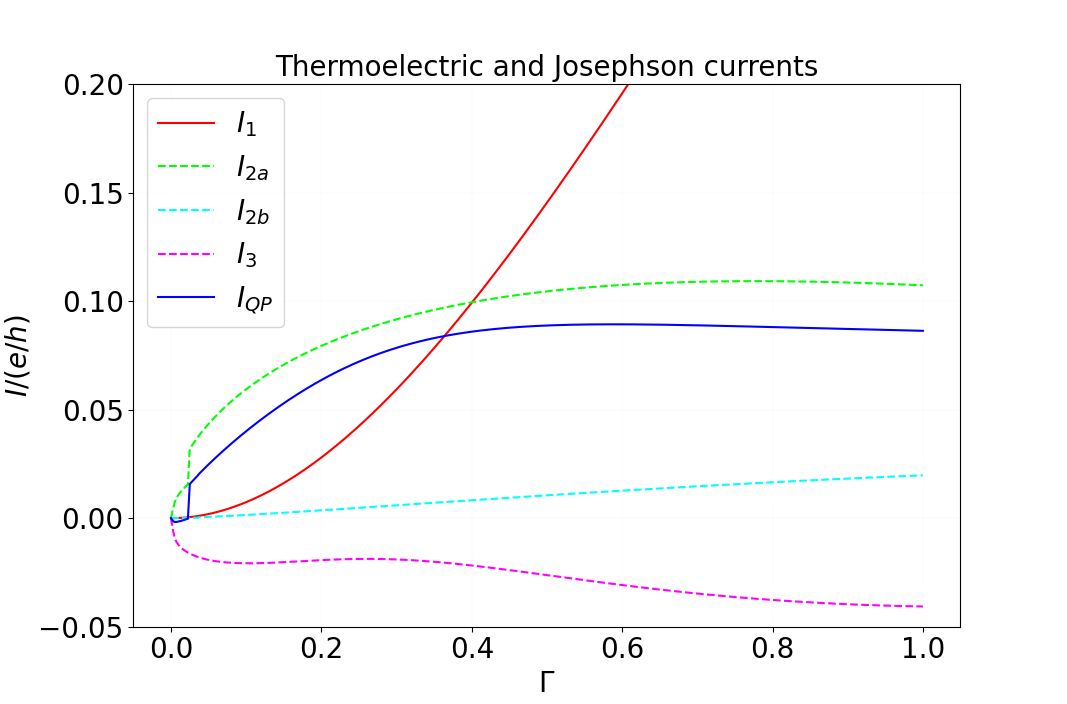}
     \caption{Currents, in units of $e/h$, for $\varepsilon_0=1$, $T_R=0.1$, $T_L=0.5$, $\Delta_R\approx 0.97$, $\Delta_L\approx 0.51$, and $\phi=\pi/2$, as functions of $\Gamma$.}
  \label{fig:currents_Gamma}
  \end{figure}

\begin{figure}[h!]
     \includegraphics[width=\linewidth]{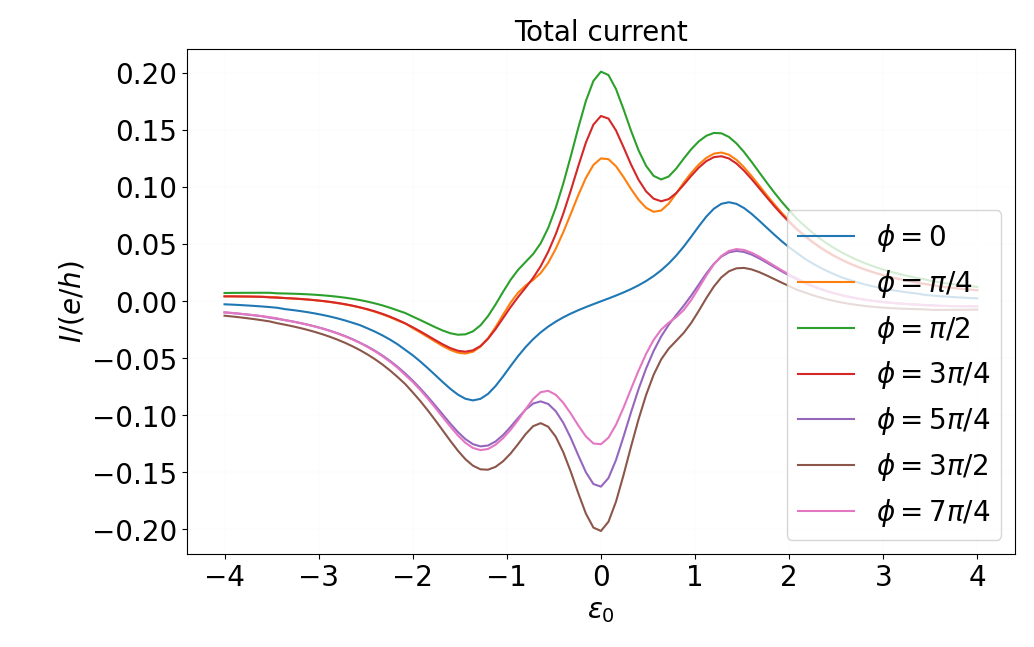}
      \caption{Total current $I$, in units of $e/h$, for $\Gamma_R=\Gamma_L=0.3$, $T_R=0.1$, $T_L=0.5$, $\Delta_R\approx 0.97$, $\Delta_L\approx 0.51$, as a function of $\varepsilon_0$, for different values of $\phi$.}
     \label{fig:currents_e0}
   \end{figure}

\begin{figure}[h!!]
    \includegraphics[width=\linewidth]{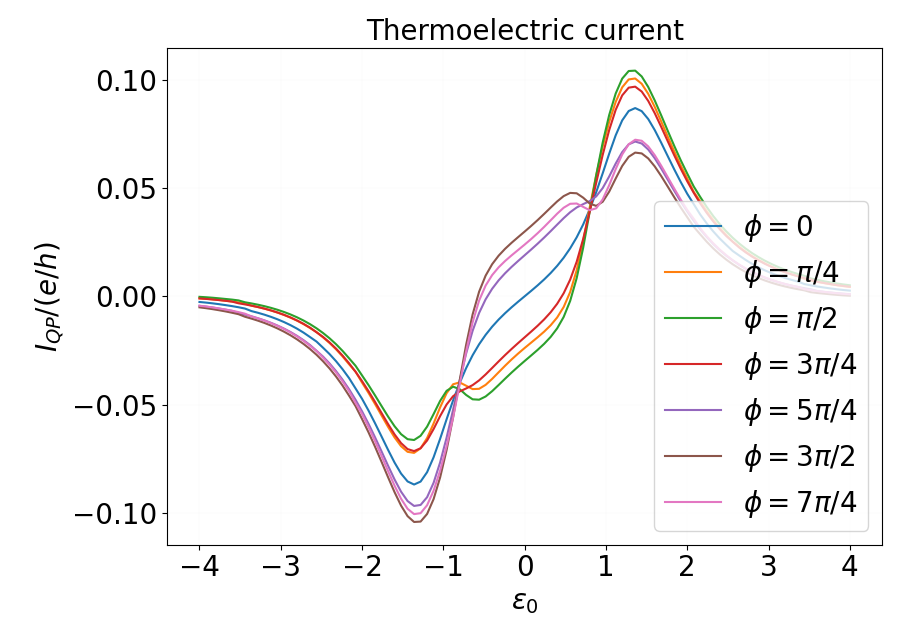}
     \caption{Thermoelectric current {$I_{QP}=I-I_1$}, in units of $e/h$, for $\Gamma_R=\Gamma_L=0.3$, $T_R=0.1$, $T_L=0.5$, $\Delta_R\approx 0.97$, $\Delta_L\approx 0.51$, as a function of $\varepsilon_0$, for different values of $\phi$.}
  \label{fig:currents_IQ_e0}
\end{figure}

If the temperatures differ consistently we have to consider different values of the gap amplitude for the right and the left superconductors. In order to have finite superconducting gaps the temperatures are bounded by the critical one. We have the following universal relations \cite{bib:tinkham}
\begin{eqnarray}
\label{delta0}
&&\Delta(T=0)\equiv\Delta_0=\alpha_o T_c\\
&&\Delta(T\lesssim 
T_c)\simeq \alpha_{c} \sqrt{T_c(T_c-T)}
\label{deltac}
\end{eqnarray}
with $\alpha_o\approx 1.764$ and $\alpha_c\approx 3.06$. 
We can adopt the following approximated form which interpolates the zero temperature and the critical temperature behavior
\begin{equation}
\label{deltaT}
\Delta(T)\approx \alpha_{c}\sqrt{T_c(T_c-T)}+(\alpha_0-\alpha_c)T_c(1-T/T_c)^\eta
\end{equation}
In order to grantee both the monotonicity and the critical asymptotic behavior, the exponent has to be 
$0.5 < \eta \lesssim 1.2$. The gap as a function of $T$ is reported in Fig.~\ref{fig:delta} with $\eta= 1$.
Now we can define $\Delta_R=\Delta(T_R)$ and $\Delta_L=\Delta(T_L)$ and calculate, for symmetric contacts, $\Gamma=\Gamma_R=\Gamma_L$, the full current
\begin{equation}
I=I_1+I_{2\rm a}+I_{2\rm b}+I_{3}
\end{equation}
from Eqs.~(\ref{eq.I1J}), (\ref{I2ab}) and (\ref{IFflast2_2}), for different sets of parameters. 
We first checked that using the same parameters of Ref.~\cite{bib:kleeorin}, we obtain the same results. 
In Fig.~\ref{fig:currents1} we report the Josephson component $I_1$ together with all the other contributions. As one can see, 
for still not too large $\delta T$, the current $I_3$ has an almost mirrored profile as compared to $I_{2\rm b}$, with a shift from zero $\propto \delta T^2$, while the main contribution to the thermoelectric transport is due to $I_{2\rm a}$, for finite values of $\varepsilon_0$. As a result the quasiparticle current {$I_{QP}=I-I_1$} is weakly dependent on the phase $\phi$, in agreement to Ref.~\cite{bib:kleeorin}.
Increasing $\delta T$, then the profiles of $I_3$ and $I_{2\rm b}$ start differing consistently, as one can see from Fig.~\ref{fig:currents_phi}, therefore, the total quasiparticle current becomes more dispersive upon varying the phase difference $\phi$. Moreover, from Fig.~\ref{fig:currents_Gamma}, we can see that, for small tunneling parameters, the thermoelectric current dominates the transport, when compared to the Josephson contribution.  
  Finally, in Fig.~\ref{fig:currents_e0} we show the total current as a function of $\varepsilon_0$ for different values of $\phi$. The current develops a peak at $\varepsilon_0=0$, for finite values of $\phi$, due to the resonance condition of the Josephson current, while the other features are due to the purely thermoelectric contributions which are singled out in Fig.~\ref{fig:currents_IQ_e0}.
  From Fig.~\ref{fig:currents_IQ_e0}, 
  we observe that, contrary to what happens in linear response regime, where thermoelectric effects appear only in the presence of an explicit breaking of the electron-hole symmetry, namely for $\varepsilon_0\neq 0$ (see Eq.~(\ref{current_LR}), since $I_{2\rm a}=0$ for $\varepsilon_0=0$), 
far from linear response regime, thermoelectric effects survive also in the presence of particle-hole symmetry, $\varepsilon_0=0$, providing that $\phi\neq 0$, in agreement with the discussion reported  Ref.~\cite{bib:marchegiani}. 
Moreover, quite intriguingly, looking at Fig.~\ref{fig:currents_IQ_e0}, we observe two nodes of the thermoelectric current as a function of the dot energy level, where the current becomes almost independent on the phase difference $\phi$.  

\section{Currents in S-QD-S Junctions with Strong Coulomb Interaction}
\label{sec:SQDScoulomb}
{\color{black}
There is an interesting case for which the current through a S-QD-S junction is time-independent even if $V\not=0$. This is the case of a strongly interacting {\color{black} single-level} quantum dot with Coulomb repulsion
\begin{equation}
\label{U}
    \hat{U}=U \hat{n}_{0\uparrow}\hat{n}_{0\downarrow}
\end{equation}
with $\hat{n}_{0\sigma}=\hat{d}^\dagger_{0\sigma}\hat{d}_{0\sigma}$.  
When $U\gg \Delta_L,\Delta_R\gg \Gamma_L,\Gamma_R$, corresponding to the experimental regime considered in Ref.~\cite{bib:ralph}, 
double occupancy of the dot levels becomes very unlikely and Andreev processes 
are strongly suppressed, thus the current is mainly produced by single quasi-particle processes ~\cite{bib:kang}. This can be explained by the fact that strong interaction strongly polarized the dot where, therefore, superconductivity is consequently suppressed.
In this case, applying our approach, we recover exactly the same results reported in Ref. \cite{bib:levyyeyati2}. 
Moreover, we calculated the thermoelectric coefficient. All the details can be found in Appendix \ref{app:SQDScoulomb}.
}
\section{Currents in N-QD-S Junctions}
\label{sec:NQDS}
{\color{black}
Another paradigmatic example of non-equilibrium problem where to test our KFT approach is N-QD-S junction. 
Also in this case we successfully benchmark the result for the current and the conductance with those already present in literature \cite{bib:sun, bib:cuevas}. In this case the current has a simple form because, due to stationarity condition, its expression does not contain the anomalous terms. Nevertheless the normal Green' s functions depend on the BCS quasi-particle DOS, therefore, we can have several processes: electron transfer, Cooper pair creation, particle-hole conversion (Andreev transmission), and Andreev reflections. Detailed calculations for the current, conductance and thermoelectric coefficient are reported in Appendix \ref{app:NQDS}.
}

\section{Conclusions}
We developed a Keldysh field theory technique to investigate the transport properties of a quantum dot contacted to superconducting reservoirs. We calculated the direct current at equilibrium and the electric and thermoelectric currents that are activated when a voltage or temperature bias drives the system out of equilibrium, also for normal-quantum dot-superconductor junction. Specifically, we derived and provided 
the explicit expression of the thermoelectric current in a superconductor-quantum dot-superconductor system. We demonstrated that, in the linear response regime, under the condition that electron-hole symmetry is violated, a thermoelectric contribution takes place in addition to the Josephson current. 
{\color {black}More intriguingly, far from linearity, we show that other contributions emerge that result in thermoelectric effects even when particle-hole symmetry is preserved. 

Our results show that S-QD-S junctions provide a promising route to achieve sizeable and tunable thermoelectric responses in superconducting systems without relying on magnetic fields or impurities. The gate-tunable energy levels of the quantum dot naturally break particle-hole symmetry, while the superconducting phase difference offers an additional knob to control both magnitude and sign of the thermoelectric current. These features make the proposed effects directly accessible with present experimental platforms where temperature gradients and phase biases can now be independently tuned. As such, our predictions offer experimentally testable signatures and a theoretical framework that can guide the design and interpretation of future thermoelectric transport experiments in hybrid superconducting devices.}

\label{sec:conclusion}

\subsection*{Acknowledgements}
The authors acknowledge financial support  
from the European Union-Next Generation EU within the ``National Center for HPC, Big Data and Quantum Computing'' (Project No. CN00000013, CN1 Spoke 10 - Quantum Computing) and from the Project "Frontiere 
Quantistiche" (Dipartimenti di Eccellenza) of the Italian Ministry for Universities and Research.

\appendix
\color{black}

\section{Currents in S-QD-S Junctions with Strong Coulomb Interaction}
\label{app:SQDScoulomb}
There exists a case for which the current through a S-QD-S junction is time-independent even if $V\not=0$. 
Let us consider 
the case of a strongly interacting {\color{black} single-level} quantum dot with Coulomb repulsion in Eq.~(\ref{U}). 
{\color{black}
As already said in the main text, when $U\gg \Delta_L,\Delta_R\gg \Gamma_L,\Gamma_R$, corresponding to the experimental regime considered in Ref.~\cite{bib:ralph}, 
double occupancy of the dot levels becomes very unlikely and Andreev processes 
are strongly suppressed, thus the current is mainly produced by single quasi-particle processes. 
} 
The condition above is, then, equivalent to neglecting terms with off-diagonal Green's functions of the coupled QD in the Nambu representation, namely {\color{black} $G^{R(A)\,12}_{00}\simeq G^{R(A)\,21}_{00}\simeq 0$, as also assumed in Ref.~\cite{bib:kang}}. {\color{black}
Under this assumption, we can perform the trace over the Nambu space in Eq.~(\ref{eq:Ia(t)SQD}) and move to frequency space, to get the stationary current
\begin{equation}
\begin{split}
\label{eq:Iqp}
   &I_a=\frac{ie}{2h}
   \int_{-\infty}^{\infty}d\omega\; \bigg\{\Gamma^{11}_{a}(\omega-\mu_a)\bigg[G_{00}^{K\,11}(\omega)+\\&-F_a(\omega-\mu_a)\big(G^{R\,11}_{00}(\omega)-G^{A\,11}_{00}(\omega)\big)\bigg]-\Gamma^{22}_{a}(\omega+\mu_a)\\
 &\times \, \bigg[G_{00}^{K\,22}(\omega)-F_a(\omega+\mu_a)\big(G^{R\,22}_{00}(\omega)-G^{A\,22}_{00}(\omega)\big)\bigg]\bigg\}
 \end{split}
\end{equation}
where also the Green's functions $G^{R(A)}(\omega)$ depend on $\mu_L$ and $\mu_R$ through the left and right self-energies.  
If we assumed spin-symmetry in the system, $G^{R11}(-\omega)=-G^{A22}(\omega)$, then Eq.~(\ref{eq:Iqp}) simplifies as follows
 \begin{equation}
\begin{split}
 I_a=&\frac{ie}{h}
 \int_{-\infty}^{\infty}d\omega\; \Gamma^{11}_{a}(\omega-\mu_a)\bigg[G_{00}^{K\,11}(\omega)+\\&-F_a(\omega-\mu_a)\big(G^{R\,11}_{00}(\omega)-G^{A\,11}_{00}(\omega)\big)\bigg],
\end{split}
\end{equation}
where, from Eq.~(\ref{gRAgeneric}), we introduced the coupling matrices
\begin{equation}
\begin{split}
    \Gamma^{11}_{a}(\omega)&=2\pi\sum_k W_{ka}^*\frac{\omega+\omega_{ka}}{2E_{ka}}\big[\delta(\omega-E_{ka})
    +\\&-\delta(\omega+E_{ka})\big]W_{ka},\\
    \Gamma^{22}_{a}(\omega)&=2\pi\sum_k W_{ka}\frac{\omega-\omega_{ka}}{2E_{ka}}\big[\delta(\omega-E_{ka})+\\&-\delta(\omega+E_{ka})\big]W_{ka}^*.
\end{split}
\end{equation}
By symmetrizing the two currents, we get the generalized Meir-Wingreen formula \cite{bib:meir,bib:wingreen,bib:kang} for the total quasi-particle current  
\begin{equation}
\label{eq:meir-wingreengen}
\begin{split}
    I=&\frac{ie}{2h}
    \int_{-\infty}^{\infty}d\omega\; \bigg[\big(\Gamma^{11}_{L}(\omega-eV)-\Gamma^{11}_{R}(\omega)\big)G_{00}^{K\,11}(\omega)\\
    &-\big(F_L(\omega-eV)\Gamma^{11}_{L}(\omega-eV)-F_R(\omega)\Gamma^{11}_{R}(\omega)\big)\\
    &\times\,\big(G^{R\,11}_{00}(\omega)-G^{A\,11}_{00}(\omega)\big)\bigg].
    \end{split}
\end{equation}
}

For a single-level dot with energy $\varepsilon_0$, the situation of a strong Coulomb interaction {with \color{black} $U\gg \Delta\gg \Gamma$} can be simply mimicked by replacing the interacting dot Hamiltonian by an effective non-degenerate single level one with energy $\varepsilon$ \cite{bib:levyyeyati2}, namely
\begin{equation}
    \hat{H}^{eff}_{dot}=\varepsilon \,\hat{d}_0^\dagger\hat{d}_0,
\end{equation}
where in $\varepsilon$ all the charging effects are included. {\color{black} Indeed, this simple yet physically intuitive assumption allows to correctly reproduce correctly the experimental results in the proper range of validity \cite{bib:ralph}.} 

In this way, the current can be written in the generalized Landauer form 
\begin{equation}
\label{eq:landauergen}
     I=\frac{e}{h}\int_{-\infty}^{\infty}d\omega\;\big[f_L(\omega-eV)-f_R(\omega)\big]\mathcal{T}(\omega,V),
\end{equation}
where the transmission coefficient is given by
\begin{equation}
    \mathcal{T}(\omega,V)=\frac{\Gamma^{11}_L(\omega-eV)\,\Gamma^{11}_R(\omega)}{(\omega-\varepsilon)^2+\frac{1}{4}\big[\Gamma^{11}_L(\omega-eV)+\Gamma^{11}_R(\omega)\big]^2},
\end{equation}
with the following analytical form for the couplings 
\begin{equation}
    \Gamma^{11}_a(\omega)=2\pi |W_a|^2\nu_a^{11}(\omega) = \Gamma_a\frac{|\omega|\theta(|\omega|-|\Delta_a|)}{\sqrt{\omega^2-|\Delta_a|^2}}, 
\end{equation}
where $\Gamma_a=2\pi|W_a|^2\nu_a$
and we considered, for simplicity, $k$-independent tunneling matrices $W_a$. Indeed, if $|\Delta_a|=0$, we exactly recover the transmission coefficient of a N-QD-N junction in the wide-band limit \cite{bib:uguccioni}. 

\begin{figure}
  \includegraphics[width=\linewidth]{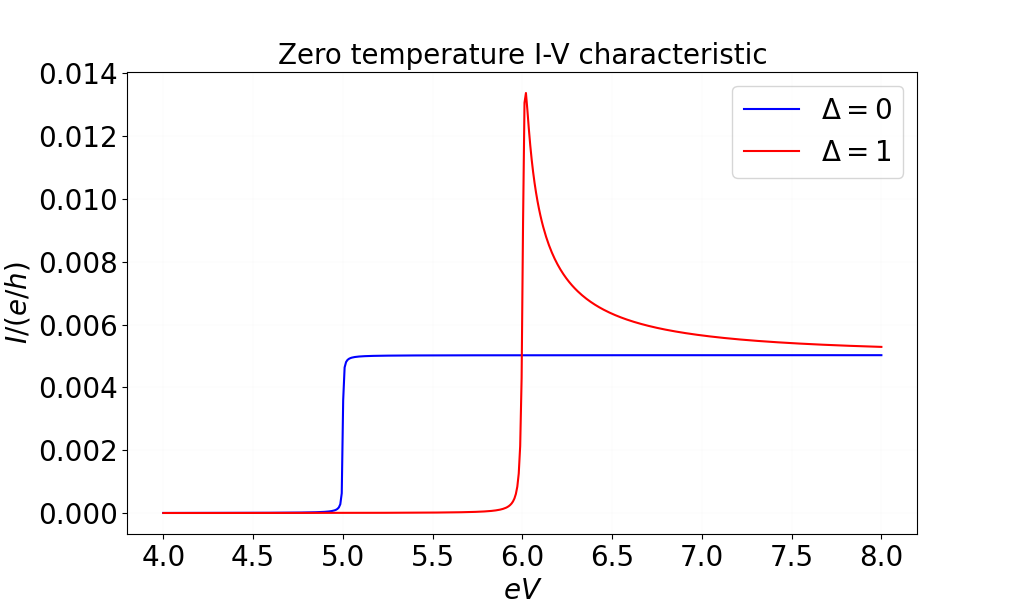}
  \caption{\color{black}Zero temperature $I-V$ characteristic in units of $e/h$ in normal ($\Delta=0$) and superconductive ($\Delta=1$) asymmetric junctions coupled, with strengths $\Gamma_L=10^{-3}$ and $\Gamma_R=4\Gamma_L$, to a strongly interacting dot with effective energy level $\varepsilon=5\equiv5\Delta$. }
  \label{fig:1} 
\end{figure}

{\color{black} 
In Fig.~\ref{fig:1} we plot the zero temperature current-voltage characteristic $I(V)$, given by 
\begin{equation}
\label{eq:I(V)zerotemp}
    I(V)=\frac{e}{h}\int_{0}^{eV}d\omega\;\mathcal{T}(\omega,V),
\end{equation}
for normal ($\abs{\Delta_L}=\abs{\Delta_R}\equiv\Delta=0$) and superconducting ($\Delta=1$) asymmetric junctions ($\Gamma_L=10^{-3}$, $\Gamma_R=4\Gamma_L$), with effective energy level $\varepsilon=5\equiv 5\Delta$, exactly reproducing the results in Ref.~\cite{bib:levyyeyati2}.}
{\color{black}Actually, in the weak coupling limit, namely for $\Gamma_L$ and $\Gamma_R$ very small, which is the limit of validity of this approach, Eq.~(\ref{eq:I(V)zerotemp}) reduces to
\begin{equation}
    I(V) \simeq\frac{2\pi e}{h}\frac{\Gamma^{11}_L(\varepsilon-eV)\Gamma^{11}_R(\varepsilon)}{\Gamma^{11}_L(\varepsilon-eV)+\Gamma^{11}_R(\varepsilon)} \,.
\end{equation}}

\begin{figure}
  \includegraphics[width=\linewidth]{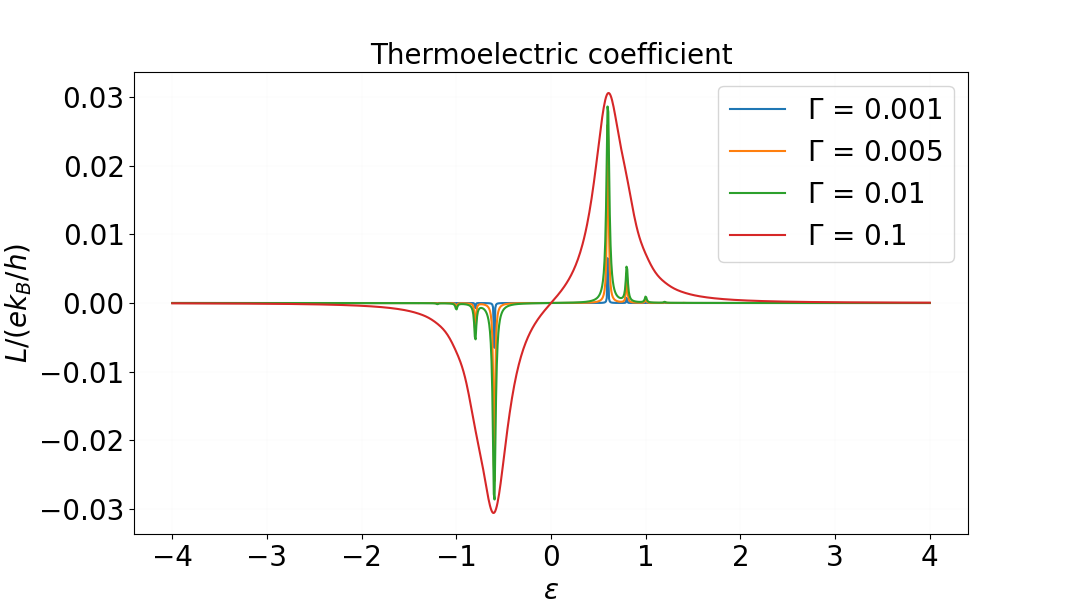}
  \caption{Thermoelectric coefficient $L$ in units of $ek_B/h$ as a function of the effective (strongly-interacting) dot's energy $\varepsilon$, in a S-QD-S junction with energy gap $\Delta=0.5$ and temperature $\beta=10$, for different values of the coupling $\Gamma$.}
  \label{fig:2}
\end{figure}

In the linear response regime, namely $V$ and $\delta T$ small (where $\delta T=T_L-T_R>0$ is the temperature difference between the leads), the  current reads $I=GV+L\delta T$, where $G$ is the linear conductance and $L$ is the {thermoelectric coefficient}. In this regime, for leads made by the same material, $|\Delta_L|=|\Delta_R|\equiv\Delta$, we have $\Delta(T_L)\simeq\Delta(T_R)$, and we find 
\begin{equation}
\begin{split}
\label{eq:GNQDN}
    G&=-\frac{e^2}{h}\int_{-\infty}^{\infty}d\omega\;\frac{df_L}{d\omega}(\omega)\,\mathcal{T}(\omega,0)\\&=-\frac{e^2}{h}
    \int_{-\infty}^{\infty}d\omega\;\frac{\frac{df_L}{d\omega}(\omega)\,
    \Gamma_L\Gamma_R\theta(|\omega|-\Delta)}{(\omega-\varepsilon)^2(1-\Delta^2/\omega^2)+(\Gamma_L+\Gamma_R)^2/4}
\end{split}
\end{equation}
\begin{equation}
\begin{split}
    L&=-\frac{e}{h}\int_{-\infty}^{\infty}d\omega\,\beta\omega\,\frac{df_L}{d\omega}(\omega)\,\mathcal{T}(\omega,0)=\\&-\frac{e}{h}
    \int_{-\infty}^{\infty}d\omega\;\frac{\beta\omega\frac{df_L}{d\omega}(\omega)\,
    \Gamma_L\Gamma_R\theta(|\omega|-\Delta)}{(\omega-\varepsilon)^2(1-\Delta^2/\omega^2)+(\Gamma_L+\Gamma_R)^2/4}.
\end{split}
\end{equation}
where we put $k_B=1$. 
{\color{black}The zero temperature linear conductance is strictly zero for $\Delta\not=0$, from Eq. (\ref{eq:GNQDN}), therefore, at finite temperature, the linear conductance will differ from zero only because of thermal effect which are generally small, i.e. $G/G_0 \approx 10^{-3}$ for $\Delta=0.5$ and $\beta=10$, as one can  check numerically. As also discussed in our previous work \cite
{bib:uguccioni}, we see that the thermoelectric coefficient is an odd function of the dot's energy $\varepsilon$, and is zero when we have particle-hole symmetry in the system, i.e. $\varepsilon=0$. This can be shown in Fig. \ref{fig:2} where we plot the thermoelectric coefficient as a function of $\varepsilon$ for different values of the coupling $\Gamma$, with $\Delta=0.5$ and $\beta=10$. }

\section{Currents in N-QD-S Junctions}
\label{app:NQDS}
Compared to the S-QD-S junction with $V\not=0$, this case is a simpler non-equilibrium problem due to the absence of the AC Josephson effect. Without loss of generality, in the following we will consider $\Delta_L=0$
and $\Delta_R=\Delta$ real. In this case, since the left lead is a normal metal, it is convenient to evaluate the left current rather than symmetrizing the result, which is identical since we are in a stationary state $I_L=-I_R=I$. This current reads
\begin{equation}
\begin{split}
   I&=\frac{ie}{h}\sum_{nm}\int_{-\infty}^{\infty}d\omega\; \Gamma^{11}_{L,nm}(\omega-eV)\bigg[G_{mn}^{K\,11}(\omega)+\\&-F_L(\omega-eV)\big(G^{R\,11}_{mn}(\omega)-G^{A\,11}_{mn}(\omega)\big)\bigg],
\end{split}
\end{equation}
{\color{black} 
where} the Green's functions contain not only the single quasi-particle contribution, but also the Andreev processes, which we are going to analyze below.

To see these new effects, we consider the usual single-level non-interacting dot with energy $\varepsilon_0$, and with $k$-independent tunneling matrices $W_L, W_R$. By using the Dyson equation in Nambu space for $\hat{G}_{00}^{R(A)}$ and $\hat{G}_{00}^K$, we get
\begin{equation}
\begin{split}
    G^{K\,11}_{00}&=\Sigma^{K\,11}|G^{R\,11}_{00}|^2+G_{00}^{R\,12}\Sigma^{K\,21}[G_{00}^{R\,11}]^*+\\&+G_{00}^{R\,11}\Sigma^{K\,12}[G_{00}^{R\,12}]^*+\Sigma^{K\,22}|G_{00}^{R\,12}|^2,
\end{split}
\end{equation}
with
\begin{equation}
\hat{\Sigma}^K=\hat{\Sigma}_L^K+\hat{\Sigma}_R^K=F_L\,\big(\hat{\Sigma}^{R}_L-\hat{\Sigma}^{A}_L\big)+F_R\,\big(\hat{\Sigma}^{R}_R-\hat{\Sigma}^{A}_R\big),
\end{equation}
where
\begin{equation}
\begin{split}
        \hat{\Sigma}_L^{R}(\omega)&=\begin{pmatrix}
            |W_L|^2g^{R11}_L (\omega-eV)& 0 \\ 0 & |W_L|^2g^{R22}_L(\omega+eV)
        \end{pmatrix},\\ \hat{\Sigma}_R^{R}(\omega)&=\begin{pmatrix}
            |W_R|^2g^{R11}_R (\omega) & -(W^*_R)^2g_R^{R12} (\omega)\vspace{0.1cm}\\ -(W_R)^2g_R^{R12} (\omega) & |W_R|^2g^{R,22}_R (\omega)
        \end{pmatrix},\\
    \end{split}
\end{equation}
while
\begin{equation}
\begin{split}
G^{R\,11}_{00}(\omega)&=\big(\omega+\varepsilon_0-|W_L|^2g_L^{R22} (\omega+eV)\\
&-|W_R|^2g_R^{R22} (\omega)\big)/D^R(\omega,V),\\
 G^{R\,22}_{00}(\omega)&=\big(\omega-\varepsilon_0-|W_L|^2g_L^{R11} (\omega-eV)\\
 &-|W_R|^2g_R^{R,11} (\omega)\big)/D^R(\omega,V),\\
 G^{R\,12}_{00}(\omega)&= -(W^*_R)^2g_R^{R12} (\omega)/D^R(\omega,V), \\
 G^{R\,21}_{00}(\omega)&=  -(W_R)^2g_R^{R12} (\omega)/D^R(\omega,V),
\end{split}
\end{equation}
where we dropped the frequency and voltage dependence for convenience, and introduced the denominator
\begin{equation}
\begin{split}
    &D^R(\omega,V)=\big[\omega-\varepsilon_0-|W_L|^2g_L^{R11}(\omega-eV)-|W_R|^2g_R^{R11}(\omega)\big]\\&\times\big[\omega+\varepsilon_0-|W_L|^2g_L^{R22}(\omega+eV)-|W_R|^2g_R^{R22}(\omega) \big]\\
    &-|W_R|^4\big(g_R^{R12}(\omega)\big)^2.
\end{split}
\end{equation}
Analogously for advanced quantities.
From these equations, if $\Delta=0$, we immediately see that particles and holes are independent, namely there is no mixing between the $11$ and $22$ components in the dressed Green's function, and we recover the result obtained for normal leads \cite{bib:uguccioni}. Differently, if the right lead is superconductive, $\Delta\not=0$, particles and holes are no longer independent and therefore we expect that they can give rise to peculiar physical phenomena during the transport. Indeed, by substituting these expressions in the formula for the current, after some algebraic manipulations, we get four contributions, namely $I=I^{ee}+I^{ep}+I^{eh}+I^A$, with
\begin{eqnarray}
\nonumber    I^{ee}&=\frac{8\pi^2e}{h}|W_L|^2|W_R|^2\int_{-\infty}^\infty d\omega\; \big[f_L(\omega-eV)-f_R(\omega)\big]\\&\times \,A(\omega,V)\,\nu^{11}_L(\omega-eV)\nu_R^{11}(\omega),\\
\nonumber    I^{ep}&=\frac{16\pi^2e}{h}|W_L|^2|W_R|^2\int_{-\infty}^\infty d\omega\; \big[f_L(\omega-eV)-f_R(\omega)\big]\\&\times \,B(\omega,V)\,\nu^{11}_L(\omega-eV)\nu^{12}_{R}(\omega),\\
\nonumber    I^{eh}&=\frac{8\pi^2e}{h}|W_L|^2|W_R|^2\int_{-\infty}^\infty d\omega\;  \big[f_L(\omega-eV)-f_R(\omega)\big]\\&\times \,C(\omega,V)\,\nu^{11}_L(\omega-eV)\nu^{22}_{R}(\omega),\\
\nonumber    I^{A}&=\frac{8\pi^2e}{h}|W_L|^4\int_{-\infty}^\infty d\omega\; \big[f_L(\omega-eV)-f_L(\omega+eV)\big]\\&\times \,C(\omega,V)\,\nu^{11}_L(\omega-eV)\nu^{22}_L(\omega+eV),
\end{eqnarray}
where the frequency and voltage dependent factors in each contribution are given by
\begin{eqnarray}
        A(\omega,V)&=&\abs{G_{00}^{R11}(\omega)}^2 ,\\
        B(\omega,V)&=&-\Re\Big[\big(G_{00}^{R11}(\omega)\big)^*G_{00}^{R12}(\omega)\,W_{R}^2/|W_R|^2\Big] , \\
        C(\omega,V)&=&\abs{G_{00}^{R12}(\omega)}^2 ,
\end{eqnarray}
and the DOS matrices introduced in Eq.~(\ref{eq:DOSmatrix}). Written in this form, each contribution has a clear interpretation in terms of elementary processes that can be identified by inspection of the DOS matrices $\nu^{ij}_a$. Notice that this expression is written in terms of particles (i.e. $11$ components in Nambu space) because of the spin symmetry assumption, but we must keep in mind that holes undergo complementary physical processes. Thus, $I^{ee}$ corresponds to normal electron transfer between the electrodes, indeed is the only term that survives in the $\Delta=0$ case. Similarly, $I^{ep}$ also corresponds to a net transfer of a single electron, but with a Cooper pair creation-annihilation as an intermediate state. On the other hand, $I^{eh}$ arises from processes where electrons in the normal electrode are converted into holes in the superconducting side, also known in literature as {Andreev transmission} \cite{bib:andreev}. Finally, $I^A$ arises from the {Andreev reflection} (AR) process \cite{bib:andreev}, in which a particle is transmitted from the left to the right electrode with a hole retro-reflecting backwards into the normal electrode, while a Cooper pair is created in the superconducting side. {\color{black} These results are compatible with the literature \cite{bib:sun}}.

\begin{figure}
  \includegraphics[width=\linewidth]{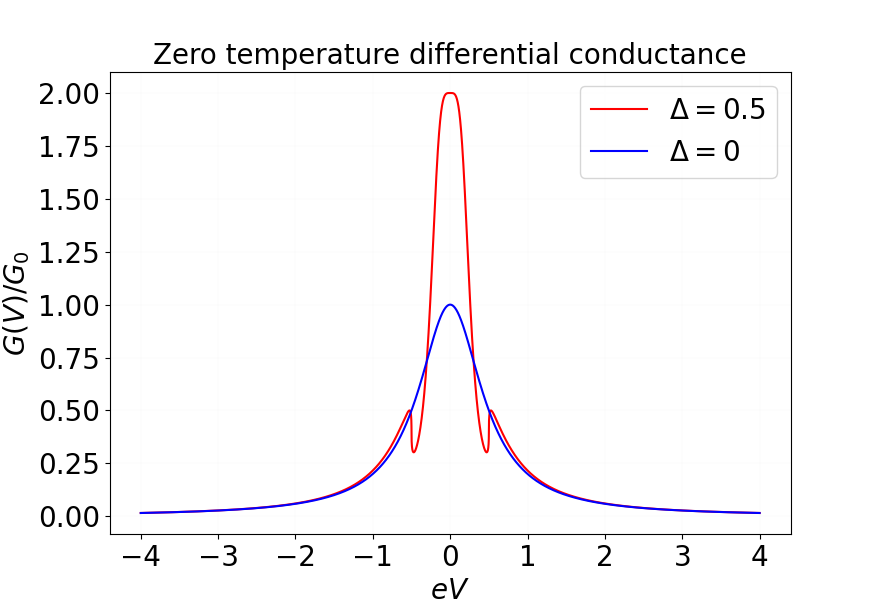}
  \caption{Zero temperature non-linear conductance in units of $G_0=2e^2/h$ as a function of the potential energy $eV$ in N-QD-N ($\Delta=0$) and N-QD-S ($\Delta=0.5$) junctions with coupling strength $\Gamma=0.5$ and dot's energy $\varepsilon_0=0$.}
  \label{fig:3}
  \end{figure}

\begin{figure}
  \includegraphics[width=\linewidth]{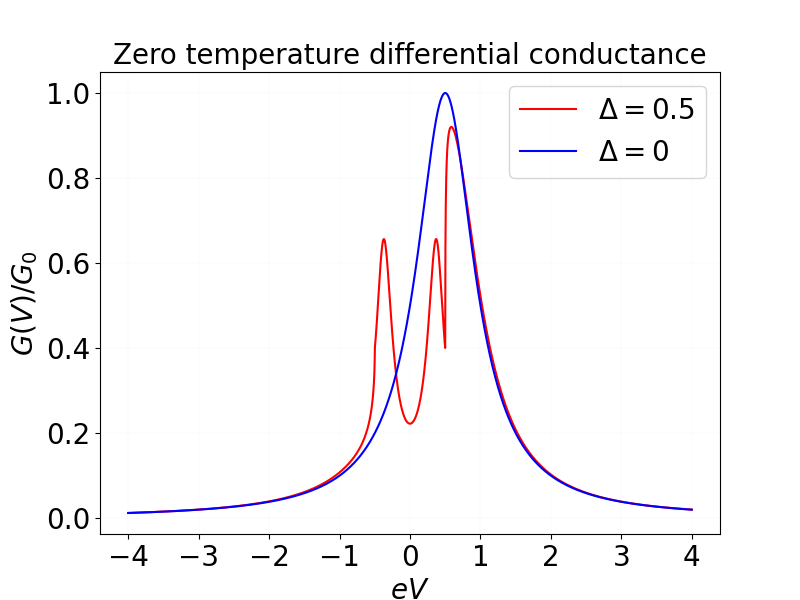}
  \caption{Zero temperature non-linear conductance in units of $G_0=2e^2/h$ as a function of the potential energy $eV$ in N-QD-N ($\Delta=0$) and N-QD-S ($\Delta=0.5$) junctions with coupling strength $\Gamma=0.5$ and dot's energy $\varepsilon_0=0.5\equiv\Delta$.}
  \label{fig:4}
\end{figure}

The total current can also be written in a simpler form, by grouping together the first three terms, getting
\begin{equation}
\begin{split}
    I&=\frac{2e}{h}\int_{-\infty}^{\infty}d\omega\; \bigg\{\big[f_L(\omega-eV)-f_R(\omega)\big]\mathcal{T}(\omega,V)\\&+\big[f_L(\omega-eV)-f_L(\omega+eV)\big]\mathcal{R}(\omega,V)\bigg\},
\end{split}
\end{equation}
where we introduced the transmission coefficient $\mathcal{T}$, corresponding to transfer processes of electronic charges, and the reflection coefficient $\mathcal{R}$, corresponding to AR processes, which is clearly zero if $\Delta=0$. By using the quasi-classical Green's functions in Eq.~(\ref{eq:gBCS}), one can  see that the transmission and reflection coefficients become voltage independent and read
\begin{equation}
\mathcal{T}(\omega)= \Gamma_L\Gamma_R\frac{A(\omega)-2(\Delta/\omega)B(\omega)+C(\omega)}{\sqrt{1-(\Delta/\omega)^2}}\theta(|\omega|-\Delta),
\end{equation}
\begin{equation}
    \mathcal{R}(\omega)=\Gamma_L^2C(\omega),
\end{equation}
where, explicitly
\begin{equation}
\begin{split}
    A(\omega)&=\bigg|\omega+\varepsilon_0+\frac{\Gamma_R\,(\omega+i0^+)}{2\sqrt{\Delta^2-(\omega+i0)^2}}+i\frac{\Gamma_L}{2}\bigg|^2 \frac{1}{|D^R(\omega)|^2},\\
    B(\omega)&=\Re\bigg[\bigg(\omega+\varepsilon_0+\frac{\Gamma_R\,(\omega+i0^+)}{2\sqrt{\Delta^2-(\omega+i0)^2}}+i\frac{\Gamma_L}{2}\bigg)^*
    \\&\times\,\frac{\Gamma_R\Delta}{2\sqrt{\Delta^2-(\omega+i0)^2}}\bigg]\frac{1}{|D^R(\omega)|^2},\\
    C(\omega)&=\bigg|\frac{\Gamma_R\Delta}{2\sqrt{\Delta^2-(\omega+i0)^2}}\bigg|^2
    \frac{1}{|D^R(\omega)|^2},\\
\end{split}
\end{equation}
and
\begin{equation}
    \begin{split}
    D^R(\omega)&=\bigg(\omega+\frac{\Gamma_R\,(\omega+i0^+)}{2\sqrt{\Delta^2-(\omega+i0^+)^2}}+i\frac{\Gamma_L}{2}\bigg)^2\\
    &-\varepsilon_0^2-\frac{\Gamma_R^2}{4}\frac{\Delta^2}{\Delta^2-(\omega+i0^+)^2}.
\end{split}
\end{equation}
At zero temperature one has
\begin{equation}
    I=\frac{2e^2}{h}\int^{V}_{0}dV'\;\big[\mathcal{T}(eV')+2\mathcal{R}(eV')\big]\equiv \int_0^V dV'\,G(V'),
\end{equation}
where the non-linear differential conductance is given by
\begin{equation}
\begin{split}
\frac{G(V)}{G_0}&=\Gamma_L\Gamma_R\frac{A(eV)-2[\Delta/(eV)]B(eV)+C(eV)}{\sqrt{1-[\Delta/(eV)]^2}}\theta(|eV|-\Delta)\\&+2{\Gamma_L^2}C(eV),
\end{split}
\end{equation}
being $G_0=2e^2/h$ the conductance quantum. We thus notice that, at zero temperature, the only term in the current which survives at $|eV|<\Delta$ is the Andreev contribution $I^A$, since it is the only process involving a reflection of quasi-particles, rather than a direct transmission, which requires a potential energy at least greater than the energy gap. Therefore, differently from the case discussed in the previous section, where, indeed, we neglected the AR processes, here we have a finite zero-temperature linear conductance, i.e. $G=G(V=0)$, given by the analytical formula
\begin{equation}
    G=2 G_0\bigg(\frac{2\Gamma_L\Gamma_R}{4\varepsilon_0^2+\Gamma_L^2+\Gamma_R^2}\bigg)^2,
\end{equation}
which becomes resonant at $\varepsilon_0=0$, and reaches the maximum $G=2G_0$, i.e. twice the conductance quantum due to tunneling of Cooper pairs, for a symmetric junction $\Gamma_L=\Gamma_R\equiv \Gamma$. 
\color{black} One can also evaluate the analytical expression for the zero temperature differential conductance evaluated at the energy gap $|eV|=\Delta\not=0$, getting
\begin{equation}
    G_\Delta=2G_0\frac{\Gamma^2_L}{4\Delta^2+\Gamma_L^2},
\end{equation}
which only comes from the AR processes, namely $G_\Delta= 2{G_0}\mathcal{R}(\Delta)$, as one can see from the fact that only the left coupling $\Gamma_L$ enters the expression (assuming that $\Gamma_R\not=0$). In particular in the quantum point contact (QPC) limit, i.e. $\Gamma_L\gg\Delta$, we obtain the maximum value $G_\Delta=2G_0$, compatible with the literature \cite{bib:cuevas}. It is also interesting to notice that $G_\Delta$ is independent on the dot's energy level $\varepsilon_0$. In Fig. \ref{fig:3} and \ref{fig:4} we report the zero temperature non-linear conductance for N-QD-N ($\Delta=0$) and N-QD-S ($\Delta=1$) junctions in the symmetric case ($\Gamma_L=\Gamma_R\equiv\Gamma=0.5$) for energy level $\varepsilon_0=0$ and $\varepsilon_0=0.5$, respectively. \color{black} We see that the behaviour depicted in the figures is perfectly compatible with the analytical results discussed above. In particular we notice that the conductance is always maximal at $eV=\varepsilon_0$ (see for example Fig. \ref{fig:3}), except the case in which $\varepsilon_0=\Delta$ and the condition $\Gamma\gg\Delta$ (QPC limit) is not satisfied (see Fig. \ref{fig:4}). \color{black}


\begin{figure}
  \includegraphics[width=\linewidth]{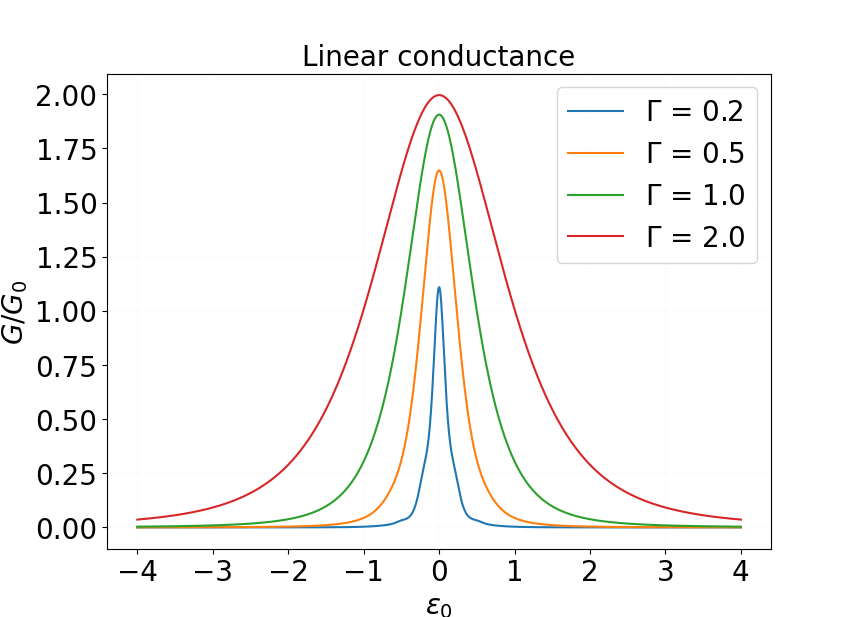}
  \caption{Linear conductance $G$ in units of $G_0=2e^2/h$ as a function of the dot's energy $\varepsilon_0$ in a N-QD-S junction with energy gap $\Delta=0.5$ and temperature $\beta=10$, for different values of the coupling $\Gamma$.}
  \label{fig:5}
\end{figure}

\begin{figure}
  \includegraphics[width=\linewidth]{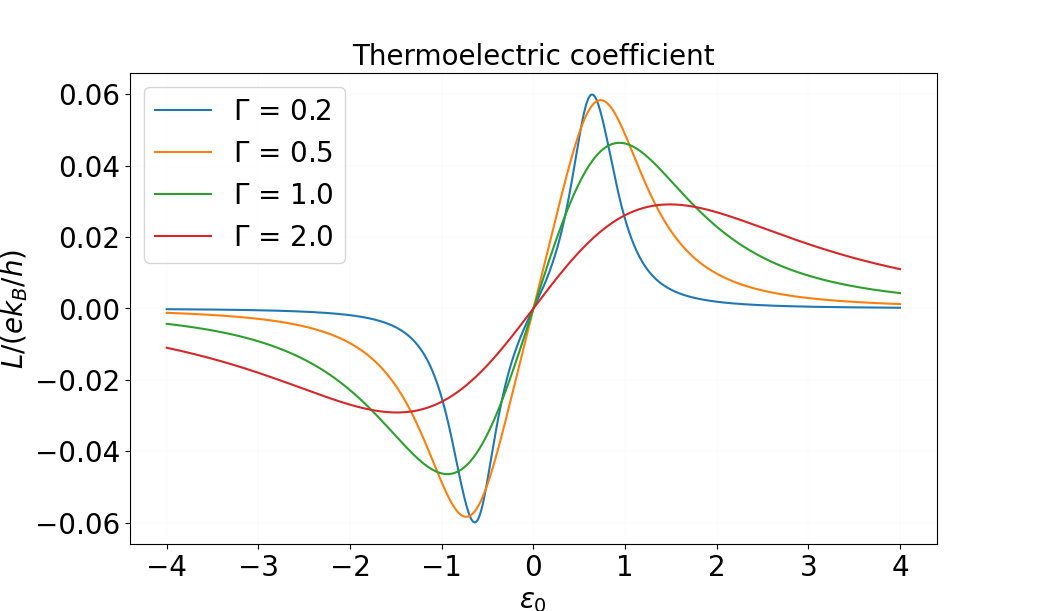}
  \caption{Thermoelectric coefficient $L$ in units of $ek_B/h$ as a function of the dot's energy $\varepsilon_0$ in a N-QD-S junction with energy gap $\Delta=0.5$ and temperature $\beta=10$, for different values of the coupling $\Gamma$.}
  \label{fig:6}
\end{figure}

At finite temperature, the linear conductance is given by
\begin{equation}
\begin{split}
    G&=-\frac{2e^2}{h}\int_{-\infty}^{\infty}d\omega\;\frac{df_L}{d\omega}(\omega)\big(\mathcal{T}(\omega)+2\,\mathcal{R}(\omega)\big)\\&\simeq-\frac{4e^2}{h}\int_{-\infty}^{\infty}d\omega\;\frac{df_L}{d\omega}(\omega)\,\mathcal{R}(\omega),
\end{split}
\end{equation}
and is shown in Fig.~\ref{fig:5} as a function of the dot's energy $\varepsilon_0$, at fixed temperature ($\beta=10$) and energy gap ($\Delta=0.5$), for different values of the symmetric coupling $\Gamma^L=\Gamma^R=\Gamma$. One can  check that the linear conductance is mainly dominated by the AR processes, since the transmission processes are very small and present only because of thermal effects as in the previous section. The thermoelectric coefficient is given by the usual formula
\begin{equation}
\label{eq:L-NQS}
    L=-\frac{2e}{h}\int_{-\infty}^{\infty}d\omega\,\beta\omega\,\frac{df_L}{d\omega}(\omega)\,\mathcal{T}(\omega),
\end{equation}
and is independent on the reflection coefficient $\mathcal{R}$, since AR processes are not produced by thermal effects. We show the thermoelectric coefficient reported in Eq.~(\ref{eq:L-NQS}) in Fig. \ref{fig:6} for different symmetric couplings $\Gamma$, at $\Delta=0.5$ and $\beta=10$.


{\color{black} 
We can see that the thermoelectric effects presented here are one order of magnitude smaller than the ones obtained for the normal case $\Delta=0$ at the same temperature \cite{bib:uguccioni}.
This result can be explained by the fact that a small thermal bias is also responsible for breaking the Cooper pairs in the leads, and thus the greater is the energy used in the latter procedure (i.e. $\Delta$), the lesser is the energy used to create a consistent transmission of single quasi-particles in the system. }

\onecolumngrid

\end{document}